\documentclass[11pt]{article}

\usepackage[english]{babel}
\usepackage[utf8]{inputenc}
\usepackage{amsmath}
\usepackage{graphicx,geometry}
\usepackage{subfig}
\usepackage[noblocks]{authblk}
\usepackage{hyperref}
\usepackage[colorinlistoftodos]{todonotes}
\newcommand{\be}{\begin{equation}}
\newcommand{\ee}{\end{equation}}
\title{{\bf Chiral Phase Transition and Meson Melting from AdS/QCD}}
\author[1]{Sean P. Bartz}
\author[2]{Theodore Jacobson}
\affil[ ]{\emph{Dept.~of Physics and Astronomy, Macalester College, St.~Paul, MN 55105}}

\affil[1]{\href{mailto:sbartz@macalester.edu}{sbartz@macalester.edu} }
\affil[2]{\href{mailto:tjacobs1@macalester.edu}{tjacobs1@macalester.edu} }

\date{\today}
\begin{document}
\maketitle

\begin{abstract}
We investigate the in-medium behavior of mesons at finite temperature and baryon chemical potential within a soft-wall model of AdS/QCD. 
We use a quartic scalar potential to obtain the correct form of chiral symmetry breaking. At zero quark mass the chiral phase transition is second-order, becoming a crossover at physical  quark mass. At zero baryon chemical potential, we find a chiral transition temperature of 155 MeV in the chiral limit and a pseudo-transition temperature of 151 MeV at physical quark mass, consistent with lattice results. In the low-temperature limit, the second-order transition occurs at a baryon chemical potential of 566 MeV while the rapid crossover occurs at 559 MeV. 
A new parameterization of the dilaton profile results in improved meson spectra. Meson melting occurs at a lower temperature and chemical potential than the chiral phase transition, so the vector-axial vector mass splitting remains constant until the bound states melt.
\end{abstract}

\section{Introduction}

The coupling constant of quantum chromodynamics changes with energy scale, becoming large enough at hadronic energy scales that perturbation theory is inadequate.
In addition, observations of heavy-ion collisions at RHIC and the LHC show that QCD matter undergoes a phase transition to a strongly-coupled quark-gluon plasma at high temperature and density.
Much effort has been put into developing non-perturbative techniques to  explore these strongly-coupled regimes of QCD.
One such tool, the AdS/CFT correspondence, shows that the same physics can be described by a strongly-coupled $d$-dimensional Supersymmetric Yang-Mills theory and a weakly-coupled gravitational theory in $d + 1$-dimensional anti-de Sitter space \cite{Maldacena1998TheSupergravity, Witten1998Anti-deTheories,Witten1998AntiHolography}. Although this conformal field theory is not confining like QCD, the correspondence has inspired phenomenological models for non-perturbative QCD. 

Over the past decade, models known as AdS/QCD or holographic QCD have been developed to describe a variety of the features of non-perturbative QCD. In order to introduce confinement, the conformal symmetry is broken by cutting off the penetration of the fields into the bulk.
``Soft-wall'' models \cite{Karch2006LinearAdS/QCD,Erlich2005QCDHadrons} achieve linear confinement by introducing a dilaton field that gradually suppresses the action. 
A dilaton that grows quadratically in the extra dimension reproduces the expected Regge trajectories of mesons at zero temperature.
Models with modified dilaton profiles have improved meson spectra and have been used to calculate other QCD observables like form factors and decay constants \cite{kwee-lebed-pion,colangelo2008,Colangelo2009HolographicMesons,Huseynova2015Model,Cui2016Infrared-improvedMesons}. For theoretical completeness, the background fields such as the dilaton should be generated dynamically from the action \cite{Batell2008,Li2013a,DePaula2010,He2013,Bartz2014DynamicalModel,Fang2016ChiralStudy}, although such effects are not explored in this work.

An important feature of low-temperature QCD is the breaking of chiral symmetry, which accounts for the existence of pions and the large mass splitting between the light vector and axial-vector mesons.
This is encoded by the non-vanishing vacuum expectation value (VEV) of the bilinear quark operator
$
\langle\bar{q}q \rangle \neq 0.
$
There are two independent sources of chiral symmetry breaking: the non-zero quark mass $m_q$, which explicitly breaks the symmetry, and a quark condensate $\sigma$, which describes the spontaneous chiral symmetry breaking. 
At high temperature, chiral symmetry is restored as the VEV of the bilinear quark operator vanishes. Lattice QCD results suggest that at vanishing baryon chemical potential, the chiral transition occurs at a lower temperature than than the deconfinement temperature \cite{Petreczky2012LatticeTemperature}. Lattice simulations, however, have difficulty in examining non-zero baryon chemical potential, while AdS/QCD models have been extended to explore its effects.

The effects of a hot and dense medium on hadronic physics are modeled holographically by changing the anti-de Sitter metric to an AdS-black hole metric, allowing for the use of black hole thermodynamic relations \cite{Witten1998Anti-deTheories, Bardeen1973TheMechanics,Hawking1983ThermodynamicsSpace}.
AdS/CFT prescribes a method for finding a spectral function for the matter fields, which allows for the study of meson and glueball melting at finite temperature \cite{Miranda2009Black-holeModel,Miranda2010GlueballsAdS/QCD,Mamani2014VectorModel,Cui2016ThermalVacuum, Chelabi2016ChiralAdS/QCD} and chemical potential \cite{Colangelo2012In-mediumQCD,Cui2013ThermalPotentialb}. 
Recent efforts have also been made to extend this melting analysis to heavy quarkonium, a useful probe for heavy ion collisions \cite{Fadafan2012OnMedium, Braga2016HolographicMelting}.
Similar techniques have also been used to describe the chiral phase transition \cite{Fang2016ChiralStudy,Chelabi2016ChiralAdS/QCD,Colangelo2012TemperatureStudy,Fang2016ChiralAdS/QCD}.

In this paper, we study the chiral dynamics of AdS/QCD at finite temperature $T$ and baryon chemical potential $\mu$. 
We extend the work presented in \cite{Chelabi2016ChiralAdS/QCD,Colangelo2012TemperatureStudy} by incorporating a non-zero baryon chemical potential in a model with independent sources of explicit and spontaneous chiral symmetry breaking.
We find that the quark condensate vanishes as either temperature or chemical potential is increased. 
With a physical quark mass, this chiral transition is a rapid crossover, becoming a second-order phase transition in the chiral limit $m_q=0$.
We introduce a new parameterization of the dilaton that improves the resulting  scalar, vector, and axial-vector meson spectra in each of these cases.
All bound states melt before the chiral transition occurs, and the mass splitting between the vector and axial-vector spectra stays  constant for all values of $T$ and $\mu$.

\section{Meson Action and Equations of Motion}
To consider the thermodynamics of soft-wall AdS/QCD, we use a 5-dimensional black hole metric that is asymptotically anti-de Sitter 
\be
ds^2=\frac{L^2}{z^2}\left(-f(z)dt^2+dx_i^2+\frac{dz^2}{f(z)}\right).\label{metricGeneric}
\ee
The function $f(z)$ has the limiting behavior $f( 0)=1$ and $f( z_h)=0$, where $z_h$ is the location of the black hole horizon. 
The AdS radius, $L$, is set to unity throughout.
We consider different functions $f(z)$ in Sections \ref{zeromu} and \ref{finitemu}, depending on the presence or absence of a baryon chemical potential.

We consider the following action for the meson fields
\be
\mathcal{S}=\frac{1}{2k} \int d^5x \sqrt{-g} e^{-\Phi(z)}\  \mathrm{Tr}\left[|DX|^2+V_m(X)+\frac{1}{2g_5^2} \left(F_A^2+F_V^2\right)\right].\label{fullaction}
\ee
Where $k$ is a factor to make the action dimensionless, $g_5^2=12\pi^2/N_c$ and $\Phi(z)$ is the dilaton. The left- and right-handed gauge fields are written in terms of the vector and axial-vector fields. The field strength tensors are defined as
\begin{eqnarray} 
F_A^{MN}&=&\partial^MA^N-\partial^NA^M-i[A^M,A^N]\\
F_V^{MN}&=&\partial^MV^N-\partial^NV^M-i[V^M,V^N],
\end{eqnarray}
and the covariant derivative is
\be 
D_MX=\partial_MX+i[V_M,X]-i\{A_M,X\},
\ee
where the indices $M, N$ run over all five coordinates. 
The complex field $X$ is dual to the $\bar{q}q$ operator, and includes the non-trivial VEV $\chi(z)$, scalar field excitations $S(x,z)$, and the pseudoscalar field $\pi(x,z)$. In a flavor-symmetric model, the VEV breaks the $SU(N_f)_L \times SU(N_f)_R$ symmetry to $SU(N_f)_V$ \cite{Karch2006LinearAdS/QCD} and the field $X$ takes the form  
\begin{eqnarray}
X^{ab}&=&\left(\frac{\chi(z)}{2}+S^a(x,z)t^b\right)I e^{2i \pi^a(x,z) t^b} \\
\langle X \rangle &=& \frac{\chi(z)}{2}I, \label{VEV}
\end{eqnarray}
where $I$ is the $N_f\times N_f$ identity matrix and the trace over flavor matrices is $\mathrm{Tr}[ t^a t^b ]=\delta^{ab}/2$. 

Higher-order terms in the scalar potential $V_m(X)$ are necessary  to allow independent sources of explicit and spontaneous chiral symmetry breaking \cite{gherghetta-kelley}. We  consider a quartic term throughout this work, along with the mass term required by the AdS/CFT dictionary
\be
V(X)=\mathrm{Tr}\left[V_m(X)\right]=\frac{m_5^2}{2} |X|^2+v_4 |X|^4,
\ee
where $m_5^2L^2=-3$. Following the analysis in \cite{Chelabi2016ChiralAdS/QCD}, we take $v_4=8$ throughout.

\subsection{Scalar Sector}

Varying the action with respect to the scalar field $X$ yields the equation of motion for the scalar mesons
\be 
\sqrt{-g}e^{-\Phi}g^{\mu\nu} \partial_\mu \partial_\nu S(x,z) +\partial_z \left(\sqrt{-g}e^{-\Phi}g^{zz}\partial_z S(x,z)\right) + \frac{\partial V}{\partial \chi} S(x,z) =0,
\ee
where Greek indices run from 0 to 3. Separating time-like and space-like components gives
\be
\frac{1}{f(z)}\partial_t^2S(x,z)-\partial_i^2S(x,z)-z^3e^{\Phi} \partial_z \left(e^{-\Phi}\frac{f(z)}{z^3} \partial_zS(x,z) \right) 
-\frac{1}{z^2}\frac{\partial V}{\partial \chi}S(x,z)=0,
\ee
with the index $i$ running from 1 to 3. Here we take the Fourier transform of the field $S(x,z) \rightarrow \tilde{S}(q,z)$ and use Proca's equation to replace $\partial_i^2 \rightarrow -q^2$, $\partial_t^2 \rightarrow -\omega^2$, resulting in
\be
-\frac{\omega^2}{f(z)}\tilde{S}(q,z)+q^2\tilde{S}(q,z)-z^3e^{\Phi}\partial_z \left(e^{-\Phi}\frac{f(z)}{z^3} \partial_z\tilde{S}(q,z) \right) 
-\frac{1}{z^2}\frac{\partial V}{\partial \chi}\tilde{S}(q,z) =0.
\ee
We perform a decomposition
\be 
\tilde{S}(q,z)=\tilde{s}(q,z)\tilde{S}_0(q), \label{fourier}
\ee
which relates the Fourier-transformed field to the source $\tilde{S}_0$ and bulk-to-boundary propagator $\tilde{s}$. 
Simplifying and taking the three-momentum $q$ to zero yields
\be 
\tilde{s}''-\left(\frac{3f(z)-zf'(z)+zf(z)\Phi'(z)}{z \ f(z)}\right)\tilde{s}'+\left(\frac{\omega^2}{f(z)^2}
-\frac{1}{z^2f(z)}\frac{\partial V}{\partial \chi}\right)\tilde{s}=0.
\ee
Finally, we introduce the dimensionless variable $u=z/z_h$, giving
\be
\tilde{s}''-\left(\frac{3f(u)-uf'(u)+ uf(u)\Phi'(u)}{ uf(u)}\right) \tilde{s}'+\left(\frac{\omega^2z_h^2}{f(u)^2}
-\frac{1}{u^2f(u)}\frac{\partial V}{\partial \chi}\right)\tilde{s}=0,\label{finalEOM} 
\ee
where $(')$ now indicates derivatives with respect to $u$.

The pions, which correspond to the pseudoscalar field $\pi(x,z)$, have a complicated equation of motion due to coupling with the longitudinal part of the axial field \cite{bartz-pions}. For this reason, analysis of this sector is left for future work.

\subsection{Vector and Axial-Vector Sectors}

Varying the action with respect to the vector field, and using the axial gauge $V_z=0$, we obtain the equation of motion
\be 
\sqrt{-g}e^{-\Phi}\partial_\nu\partial^\nu V_\mu(x,z)+\partial_z\left(\sqrt{-g}e^{-\Phi}g^{zz}\partial_z V_\mu(x,z)\right)=0.
\ee
Applying the Fourier transform and using Proca's equation we find
\be 
-\frac{\omega^2}{f(z)}\tilde{V}_\mu(q,z)+q^2\tilde{V}_\mu(q,z)-ze^{\Phi}\partial_z\left(e^{-\Phi}\frac{f(z)}{z}\partial_z\tilde{V}_\mu(q,z)\right)=0.
\ee
Performing a decomposition similar to (\ref{fourier}), we use $\tilde{V}_\mu(q,z)=\tilde{v}(q,z)\tilde{V}^0_\mu(q)$ to obtain an equation of motion for the bulk-to-boundary propagator.
Again we take three-momentum to be zero and switch to the dimensionless variable $u$, resulting in the equation of motion
\be 
\tilde{v}''-\left(\frac{f(u)-uf'(u)+uf(u)\Phi'(u)}{uf(u)}\right)\tilde{v}'+\left(\frac{\omega^2z_h^2}{f(u)^2}\right)\tilde{v}=0. \label{vectorEOM}
\ee

The equation of motion for the axial sector is found similarly. 
The coupling of the axial field to the scalar VEV $\chi(z)$ results in an additional term that is responsible for the vector-axial mass splitting.
The equation of motion for the axial meson reads
\be 
\tilde{a}''-\left(\frac{f(u)-uf'(u)+uf(u)\Phi'(u)}{uf(u)}\right)\tilde{a}'+\left(\frac{\omega^2z_h^2}{f(u)^2}-g_5^2 \frac{\chi(u)^2}{u^2f(u)}\right)\tilde{a}=0. \label{axialEOM}
\ee

\subsection{Dilaton Parameterization}

\begin{figure}[!ht] 
\centering
\subfloat[]{
  \includegraphics[width=0.5\textwidth]{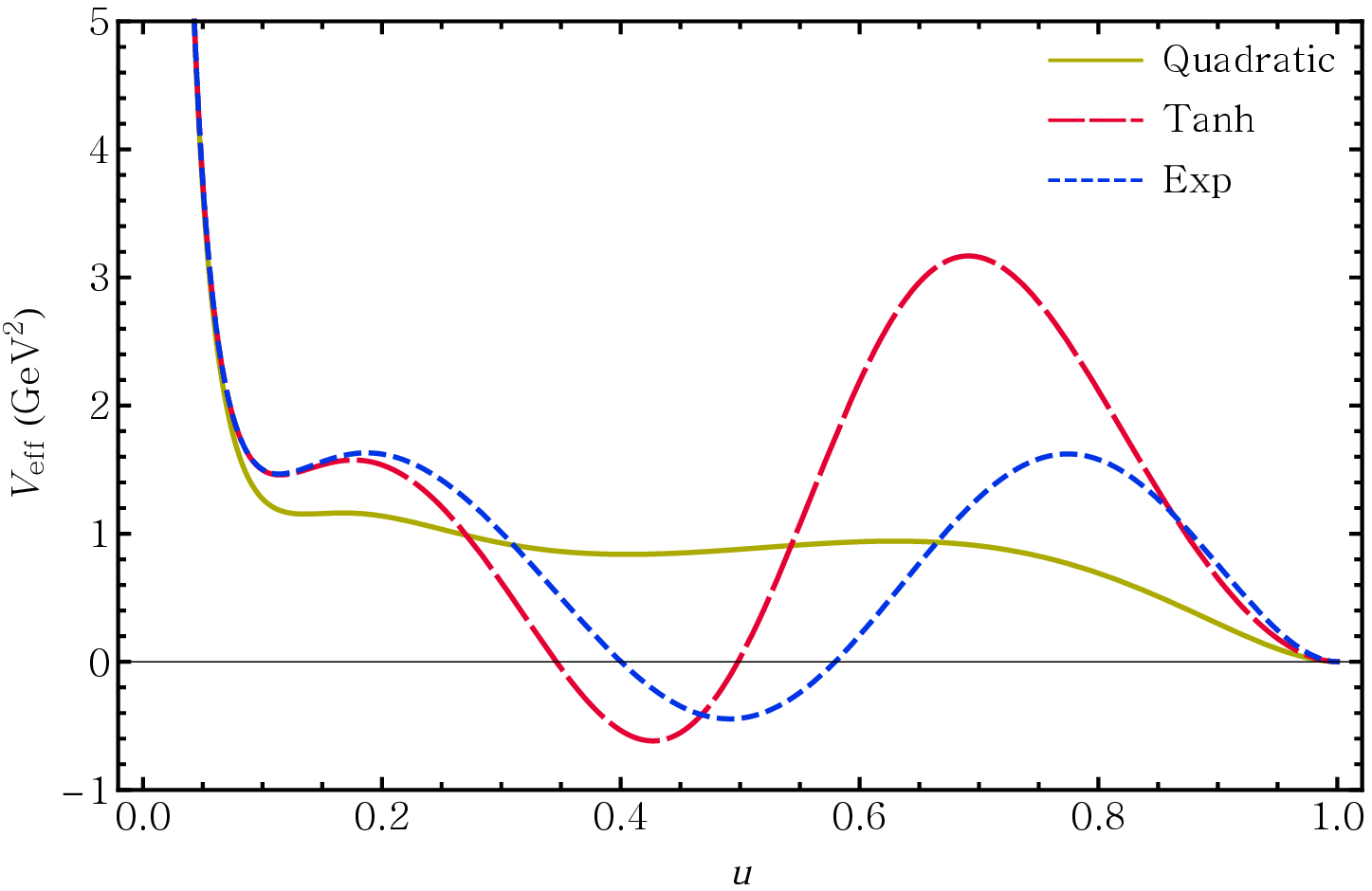}
}
\subfloat[]{
  \includegraphics[width=0.5\textwidth]{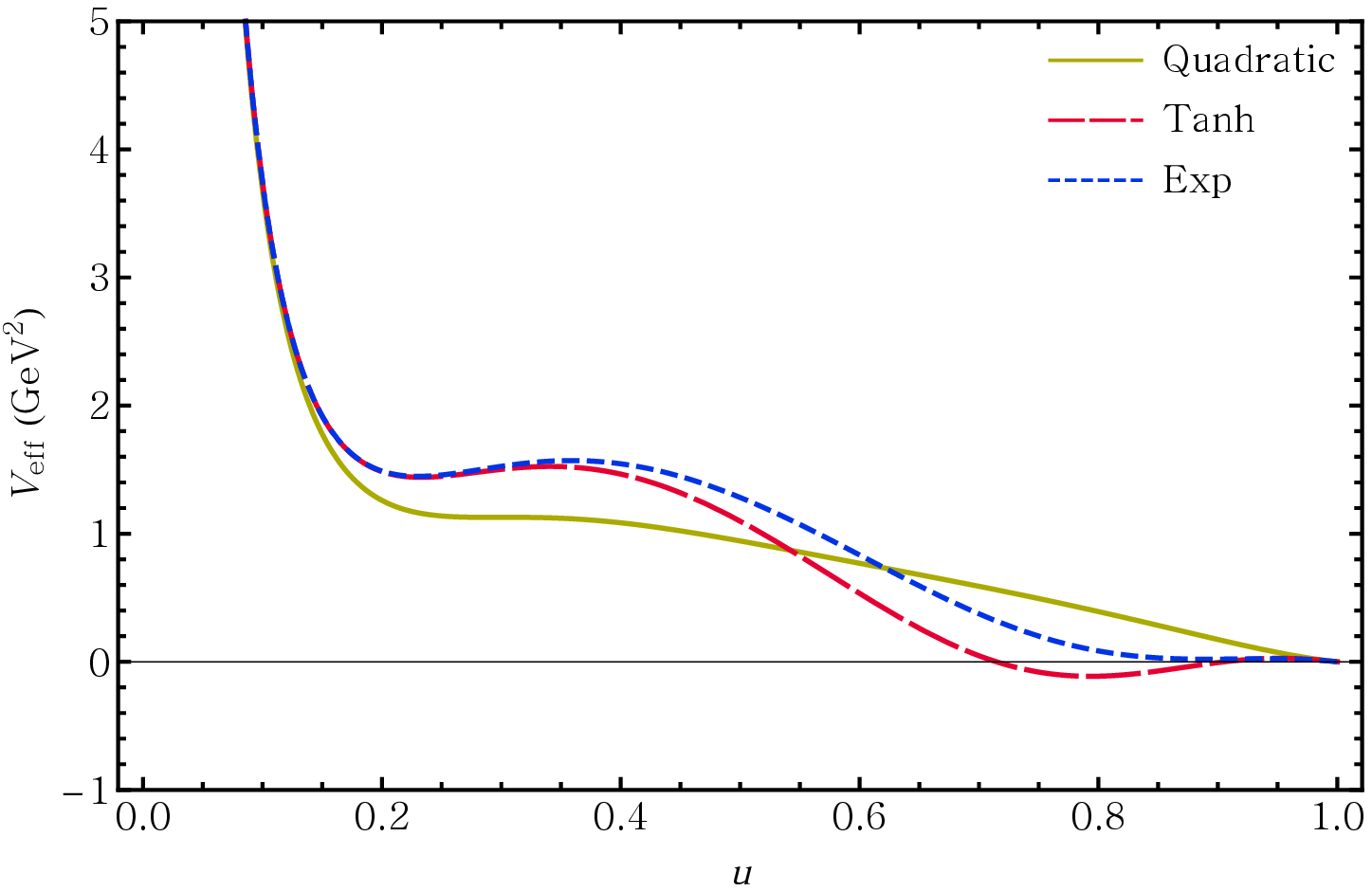}
}
\caption{Effective potential for the axial meson, with $m_q=9.75$ MeV and for two temperatures (a) $T=0.035$ GeV and (b) $T=0.070$ GeV. For different parameterizations of the dilaton, the potential differs. The exponential parameterization (\ref{expdil}) produces a potential with smaller fluctuations away from that of the quadratic dilaton, which corresponds to strictly linear meson trajectories.} \label{pot}
\end{figure}

In the soft-wall model, a quadratic dilaton profile $\Phi \sim z^2$ reproduces linear Regge trajectories \cite{Karch2006LinearAdS/QCD}, and any modification to the dilaton profile must remain quadratic in the large-$z$ (IR) limit \cite{Karch2011OnModels}.  
The authors of \cite{Chelabi2016ChiralAdS/QCD} showed that a purely quadratic dilaton profile does not allow for the dynamical solution of a chiral field when higher-order terms are included in the scalar potential. 
They find that a negative quadratic dilaton $\Phi(z) \sim -z^2$ permits a chiral field that obeys the appropriate boundary conditions, but such a dilaton profile does not produce correct spectra.
In order to obtain a correct spectrum while also dynamically generating the chiral field, they explore a dilaton parameterization that is negative quadratic in the small-$z$ (UV) limit, while becoming positive quadratic in the IR
\be 
\Phi(u)=-\mu_1^2z_h^2u^2+\left(\mu_0^2+\mu_1^2\right)z_h^2u^2\tanh\left(\mu_2^2z_h^2u^2\right). \label{tanhdil}
\ee
Based on this analysis, we choose $\mu_0=430$ MeV, $\mu_1=830$ MeV, and $\mu_2=176$ MeV. Upon inspection, we find that a more gradual transition between the UV and IR limits, given by the exponential parameterization
\be 
\Phi(u)=-\mu_1^2z_h^2u^2+\left(\mu_0^2+\mu_1^2\right)z_h^2u^2 \left[1-\exp\left(-\mu_2^2z_h^2u^2\right)\right],\label{expdil}
\ee
yields improved meson spectra. This is seen by comparing the effective Schr{\"o}dinger-like potential for each dilaton profile. The equations of motion are put into a Schr{\"o}dinger-like form through a Bogoliubov transformation of the fields. For the purposes of demonstration we choose the axial sector. The equation of motion (\ref{axialEOM}) is re-written
\be
\left(\omega^2-f(z)q^2-f(z)g_5^2\frac{\chi^2}{z^2}\right)\tilde{a}+e^{\omega_a}f(z)\partial_z\left(e^{-\omega_a}f(z)\partial_z\tilde{a}\right),
\ee
where $\omega_a \equiv \Phi(z)+\mbox{log}(z)$. Switching to the tortoise coordinate $r_*$ by using the relation $\partial_{r_*}=-f(z)\partial_z$ and applying the Bogoliubov transformation $\tilde{a}=e^{\omega_a/2}\ \psi$
yields a Schr{\"o}dinger-like equation
\be 
\partial_{r_*}^2 \psi+\omega^2 \psi = V_s\psi,\label{schro}
\ee
with the effective potential
\be 
V_s=f(z)^2\left(\frac{\omega_a'^2}{4}-\frac{\omega_a''}{2}\right)+f(z)g_5^2\frac{\chi^2}{z^2}.
\ee
Though this potential is exact only in the zero-temperature limit, we use it as an {effective} potential which we study at various temperatures. Figure \ref{pot} shows the effective potential for the quadratic, hyperbolic tangent, and exponential dilaton profiles for two temperatures. The hyperbolic tangent parameterization causes larger fluctuations in the effective potential, which raise the ground state and first excited state masses. The data in Table \ref{masstable} shows a comparison of the masses obtained by the two dilaton parameterizations for the ground state and first excited state. The exponential parameterization yields mass spectra more consistent with data, which we attribute to the smoother Schr{\"o}dinger-like potential. We choose (\ref{expdil}) as the dilaton profile for the remainder of this work.
\begin{table}[!ht]
\centering
 \begin{tabular}{c c c c c c} 
 \hline
 \hline
Meson & $n$ & Experimental (MeV) & \ $\Phi_{\exp} $ (MeV) \ & \ $\Phi_{\tanh}$ (MeV) \  \\ 
 \hline
$f_0$ & 1 & 400-550 & 776 & 905 \\ 

& 2 & 990 $\pm$ 20 & 1303  & 1439 \\
 \hline
 $\rho$ & 1 & 775.5 $\pm$ 1 & 1078 & 1174 \\ 

& 2 & 1282 $\pm$ 37 & 1428 & 1581 \\
 \hline
 $a_1$ & 1 & 1230 $\pm$ 40 & 1195 & 1265 \\ 

& 2 & 1647 $\pm$ 22  & 1510 & 1673 \\
 \hline
 \hline
\end{tabular}
\caption{Experimental \cite{Olive2014ReviewPhysics} and predicted values of meson masses. The ground state corresponds to the radial quantum number $n=1$, and the first excited state corresponds to $n=2$. Experimental data for all three sectors is shown, along with the predictions using the exponential dilaton profile and the hyperbolic tangent dilaton profile.}
\label{masstable}
\end{table}

\subsection{Method for finding the spectral function} \label{spectralmethod}
Using the Method of Frobenius, we analyze the limiting behavior of the solutions to (\ref{finalEOM}, \ref{vectorEOM}, \ref{axialEOM}) as $u\rightarrow 0$.  The leading-order power of the two independent solutions is given by the two roots $r_1,  r_2$ of the indicial equation
\be
(r-4)r-m_5^2L^2=0, \label{indicial} 
\ee
such that 
\begin{eqnarray}
\psi_1(u) &=& a_1 u^{r_1}+\cdots \\
\psi_2(u) &=& b_1 u^{r_2}+\cdots
\end{eqnarray}
It is noteworthy that requiring both roots of this equation be real is equivalent to the Breitenlohner-Freedman bound on the mass of the fields \cite{Breitenlohner1982PositiveSupergravity}, $m_5^2L^2 \geq -4$. Performing a Bogoliubov transformation on (\ref{finalEOM}) yields the Schr{\"o}dinger-like equation studied in \cite{Wolf1981Equally-spacedWell}, which finds that the condition on $m_5^2$ marks the boundary between equations with discrete and continuous mass spectra.

This analysis provides the UV boundary conditions for each sector. 
Following the AdS/CFT dictionary, we use the UV limit of the scalar VEV
\be 
\chi(u \rightarrow 0)=m_q\zeta z_h u+\frac{\sigma}{\zeta}z_h^3u^3, \label{chiralUV}
\ee
where $m_q$ is the light quark mass and $\sigma$ is the light quark chiral condensate. Large $N_c$ scaling analysis gives $\zeta=\sqrt{N_c}/2 \pi$ \cite{Cherman:2008eh}.  The two independent UV solutions for the scalar field are
\begin{eqnarray} 
 {s_1}(u) &=& a_0 \left(u^3 -m_q z_h\zeta u^4 + \frac{z_h^2}{8}\left(3m_q^2\zeta^2-6\mu_1^2-\omega^2\right) u^5\right), \label{scalar_UV} \\ 
 {s_2}(u) &=& b_0 \Bigl(u +3m_q z_h \zeta u^2-m_q z_h^3 \zeta \left(4\mu_1^2+\omega^2\right) u^4  \nonumber \\ 
 &\ & \qquad \qquad \qquad \qquad +\frac{z_h^2}{2a_0}\left(9m_q^2\zeta^2+2\mu_1^2+\omega^2\right){s_1}(u)\log u \Bigr).
\end{eqnarray}
The UV solutions to the axial field are
\begin{eqnarray} 
{a_1}(u)&=&c_0\left(u^2-\frac{4z_h^2\mu_1^2+z_h^2\omega^2-g_5^2m_q^2z_h^2\zeta^2}{8}u^4\right),\\
{a_2}(u)&=&d_0+d_2u^2-\left(\frac{-g_5^2m_qz_h^4\sigma}{4}d_0+\frac{4z_h^2\mu_1^2+z_h^2\omega^2-g_5^2m_q^2z_h^2\zeta^2}{8}d_2\right)u^4. \label{axial_UV}
\end{eqnarray}
Lastly, for the vector field the UV solutions are
\begin{eqnarray} 
{v_1}(u)&=&g_0 \left(u^2-\frac{4z_h^2\mu_1^2+z_h^2\omega^2}{8}u^4\right),\\
{v_2}(u)&=&h_0+h_2u^2-\frac{4z_h^2\mu_1^2+z_h^2\omega^2}{8}h_2u^4.
\label{vector_UV}
\end{eqnarray}
The coefficients $a_0,b_0,c_0,d_0,g_0,h_0$ provide overall normalization for the fields, and are set to unity for simplicity. 

Near the black hole horizon, the potential in (\ref{schro}) goes to zero, and the solutions  are given by the free-particle in-falling ($\psi_-$) and out-going ($\psi_+$) solutions 
\be
\psi_\pm=(1-u)^{\pm i\frac{\omega z_h}{4}}, \label{infalling}
\ee
with $\psi_-$ representing the correct choice of boundary condition.
The wavefunctions that describe the UV behavior $(\psi_1,\, \psi_2)$ and the near-horizon behavior $(\psi_-,\, \psi_+)$ form complete sets of solutions to the differential equations. Thus, we can write the near-horizon solutions as linear combinations of the UV solutions
\begin{eqnarray}
\psi_-&=&A_- \psi_1 + B_- \psi_2 \label{outgo}\\
\psi_+&=&A_+ \psi_1 + B_+ \psi_2. \label{infall}
\end{eqnarray}
Here, $\psi_1$ is chosen to be the UV solution with the larger leading-order power. When the fields are Bogoliubov-transformed, $\psi_1$ corresponds to the normalizable solution to the Schr{\"o}dinger-like problem.

In order to find the meson spectral function, we find the two-point correlator, identified as the retarded Green's function. According to the AdS/CFT dictionary \cite{Son2002Minkowski-spaceApplications}, the retarded two-point Green's function $\Pi^R$ is found by differentiating the on-shell action with respect to the source of the field. The on-shell action for the scalar field is given by 
\be
\mathcal{S}_{\mathrm{on-shell}}= \left.\lambda\int d^4x \sqrt{-g} e^{-\Phi} g^{zz}S(x,z)\partial_z S(x,z)\right|_{z=0}.
\ee
Here, we use $\lambda$ as a constant to give the correct dimensions to the on-shell action, but whose value is unimportant to the analysis. Using the Fourier transformed fields (\ref{fourier}) and separating the source from the bulk-to-boundary propagator, the action becomes
\be
\mathcal{S}_{\mathrm{on-shell}}= \left.-\lambda\int d^4x \sqrt{-g} e^{-\Phi}  g^{zz} \tilde{S}_0(-q)\tilde{s}(q,z)\partial_z \tilde{s}(q,z) \tilde{S}_0(q)\right|_{z=0}.
\ee
Replacing $\omega=q_0$ and differentiating with respect to the source, the retarded Green's function, up to an overall factor, is given by 
\be
\Pi^R(\omega^2) =  \sqrt{-g} e^{-\Phi}g^{zz}\tilde{s}(\omega^2,z)\partial_z\tilde{s}(\omega^2,z)\Big|_{z=0}.  \label{Green}
\ee
The spectral function is given by $\mathrm{Im}\, \Pi^R(\omega^2)$, and is proportional to $\mathrm{Im}\, A_-/B_-,$ which is determined by a numerical procedure outlined in Appendix \ref{numericalmethod}. Identical procedures give the spectral functions for the vector and axial-vector sectors.




\section{Finite Temperature, Zero Baryon Chemical Potential}\label{zeromu}

To consider the effects of finite temperature, we use the AdS-Schwarzschild metric by using  
\be
f(z)=1-\left(\frac{z}{z_h}\right)^4
\ee
in the black hole metric (\ref{metricGeneric}). In terms of the dimensionless coordinate, the black hole function is simply $f(u)=1-u^4$. The temperature is related to $z_h$, the position of the black hole horizon in the holographic coordinate
\be 
T=\frac{1}{4\pi}\left|\frac{df}{dz}\right|_{z=z_h}.
\ee
In the AdS-Schwarzschild metric, this reduces to $T=(\pi z_h)^{-1}$.

\subsection{Chiral Dynamics} \label{Chiral}
The chiral dynamics of the soft-wall model are described by the $z$-dependent vacuum expectation value of the scalar field (\ref{VEV}). The equation of motion for this chiral field is
\be 
\chi''(u)-\left(\frac{3f(u)-uf'(u)+uf(u)\Phi'(u)}{uf(u)}\right)\chi'(u)-\frac{1}{u^2f(u)}\frac{\partial V}{\partial \chi}=0. \label{chiralEOM}
\ee 
In the near-horizon limit, this equation has one divergent and one non-divergent solution. The non-divergent solution that we seek behaves as $\chi(u\rightarrow 1) \sim u^0.$ The UV behavior is given by (\ref{chiralUV}). 

At sufficiently low temperatures, the solutions approach a constant value that satisfies $\partial V/\partial \chi=0$.
The non-trivial solution that satisfies this condition is $\chi=\sqrt{3/4v_4}$. At higher temperatures, this limit is not reached, but a unique solution is determined numerically by requiring that the chiral field not diverge as it approaches the horizon. 
Notably, using the exponential parameterization of the dilaton (\ref{expdil}) allows us to obtain regular solutions to (\ref{chiralEOM}) at lower temperatures than when using the hyperbolic tangent parameterization (\ref{tanhdil}). 

\begin{figure}[ht] 
\centering
\subfloat[]{
  \includegraphics[width=0.5\textwidth]{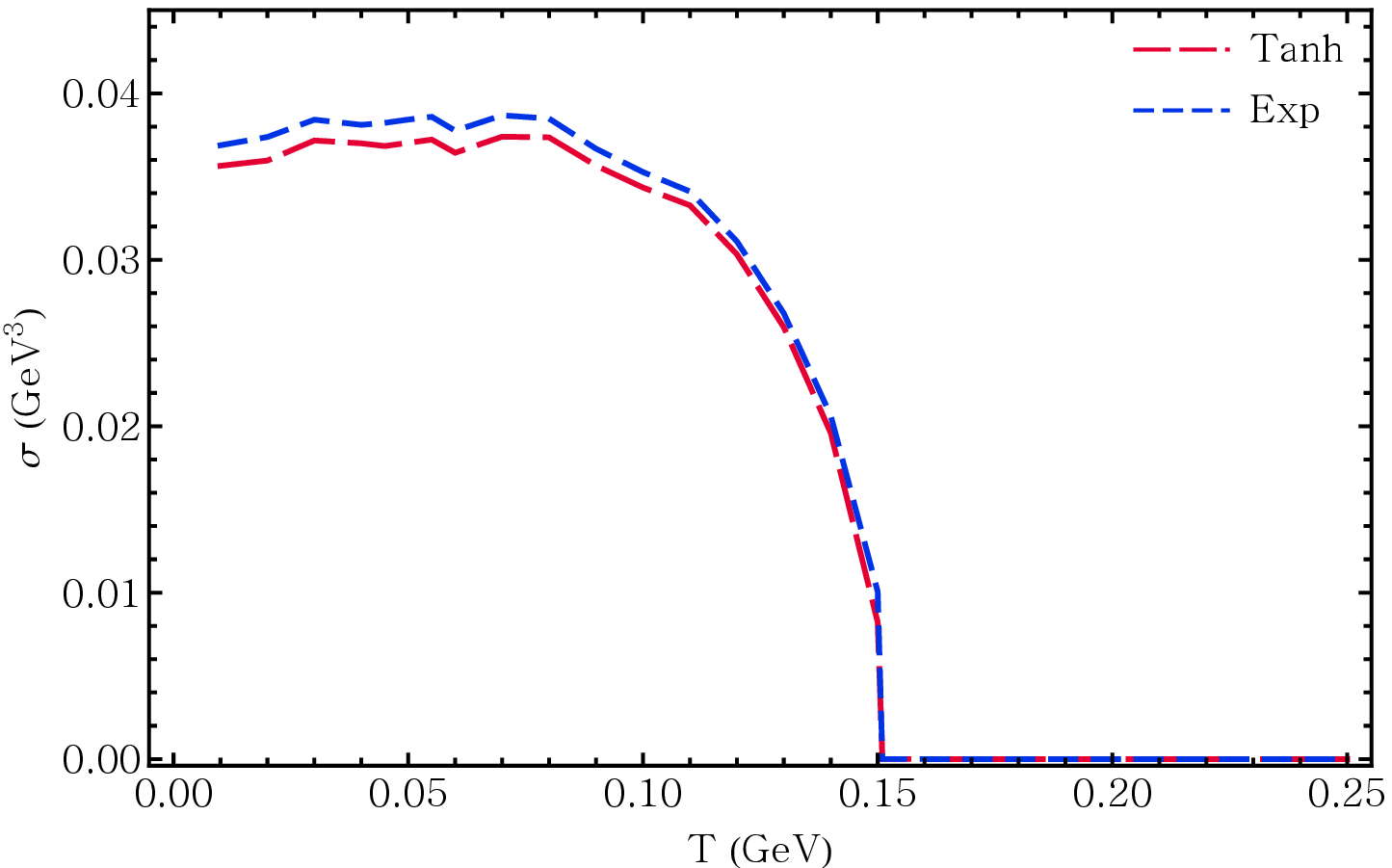}
}
\subfloat[]{
  \includegraphics[width=0.5\textwidth]{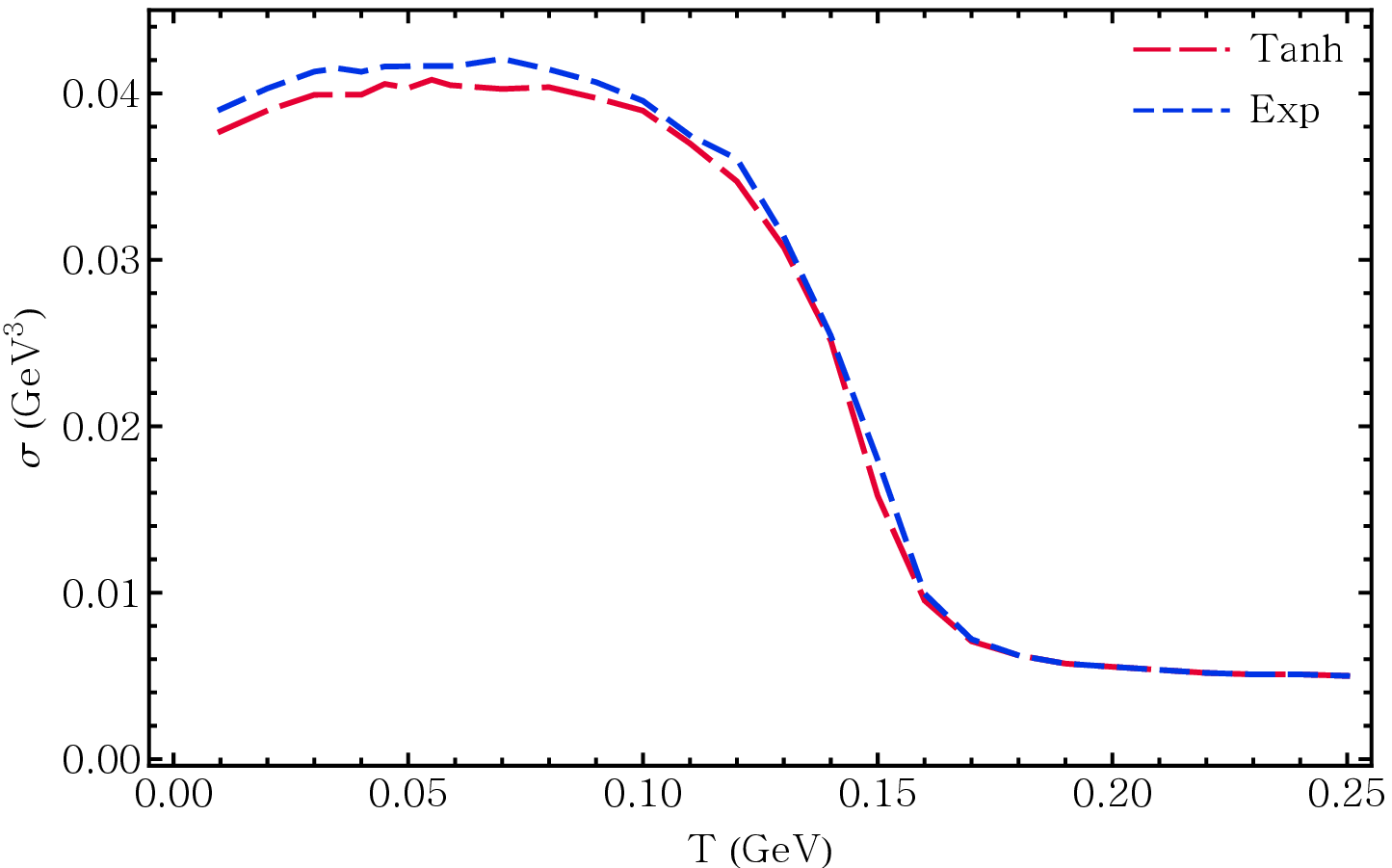}
}
\caption{Temperature dependence of $\sigma$ for both parameterizations of the dilaton, and for (a) $m_q=0$ MeV and (b) $m_q=9.75$ MeV. In each case the phase transition occurs near a temperature of $T=0.15$ GeV.} \label{sig1}
\end{figure}

We obtain the full solutions for $\chi(u)$ for each temperature using a numerical shooting method, varying $\sigma$ and finding the value that matches the boundary conditions as $u\rightarrow 1$. Because the chiral field itself is not sufficiently sensitive to variations in $\sigma$, we follow the procedure in \cite{Chelabi2016ChiralAdS/QCD}, finding the value that causes the following ``test" function to be zero as $u\rightarrow 1$
\be 
-u^2\frac{f'(u)}{f(u)}\chi'(u)-\frac{1}{f(u)}\left(3 \chi(u)-4 v_4 \chi(u)^3 \right).
\ee
The test function comprises the terms in the equation of motion (\ref{chiralEOM}) that are singular at the horizon.  We analyze the dependence of the chiral condensate $\sigma$ on temperature  in the chiral limit and for a physically reasonable value of the quark mass $m_q=9.75$ MeV. These results are shown in Figure \ref{sig1}. The chiral phase transition is higher-order in the chiral limit, becoming a crossover at the physical quark mass. The higher-order transition is not found in models lacking the nonlinear term in the scalar potential because the sources of spontaneous and explicit chiral symmetry breaking are not independent; in such models, $\sigma$ vanishes in the chiral limit. The dilaton parameterization has negligible effect on the chiral condensate.

\subsection{Meson Melting} \label{tempspectra}
\begin{figure}[ht]
  \centering
  \includegraphics[width=0.5\textwidth]{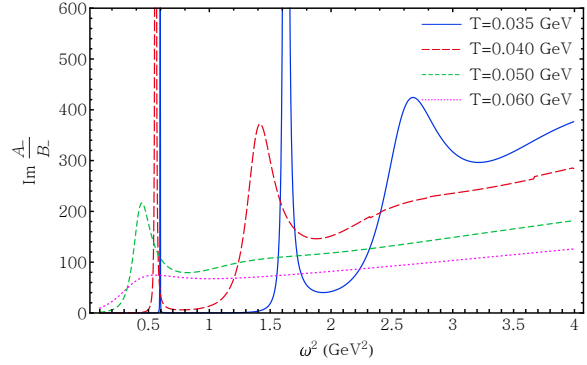}
  \caption{Thermal spectral function for scalar meson.}
  \label{scalar_spectra_t}
\end{figure}

\begin{figure}[ht] 
\centering
\subfloat[]{
  \includegraphics[width=0.5\textwidth]{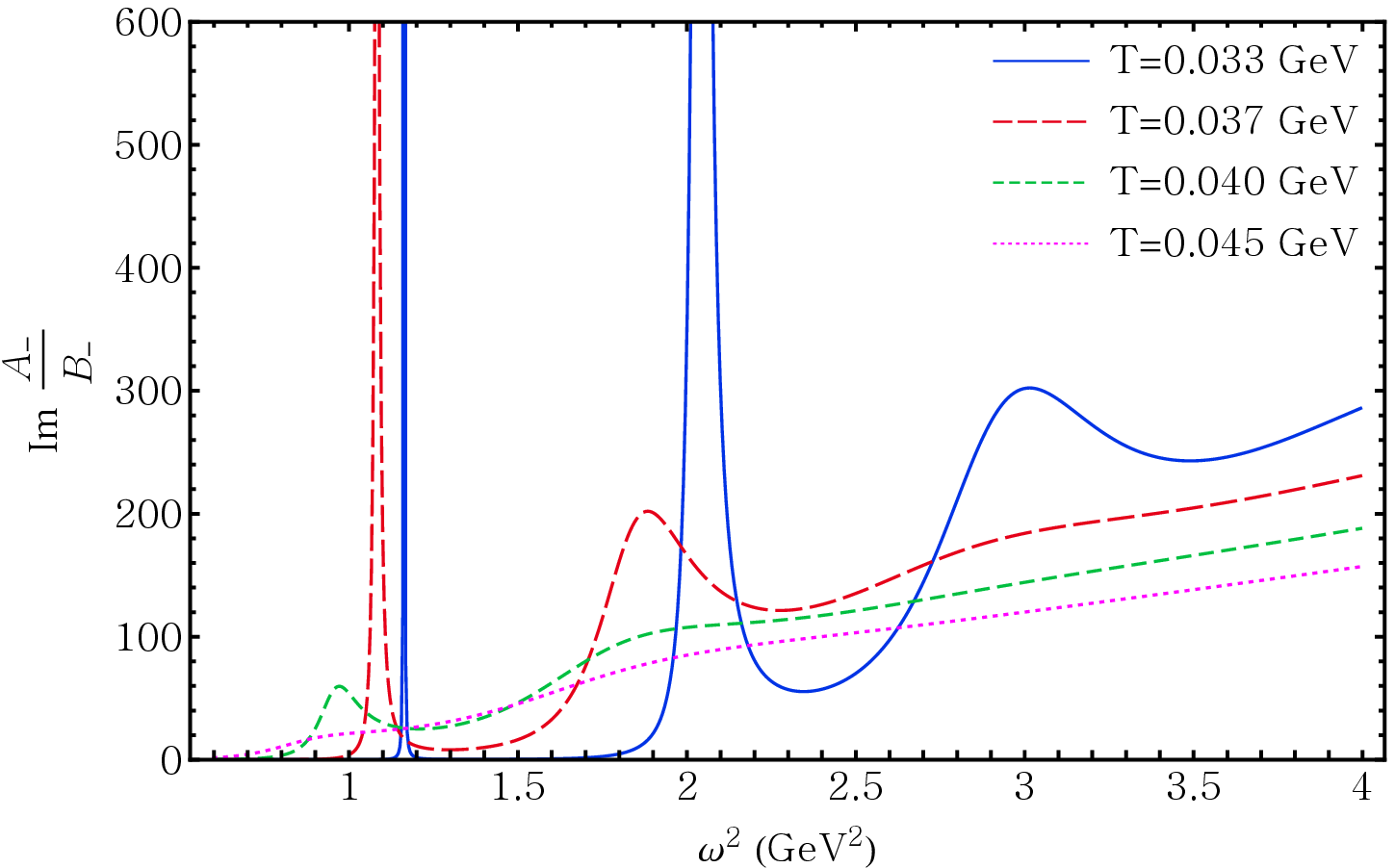}
}
\subfloat[]{
  \includegraphics[width=0.5\textwidth]{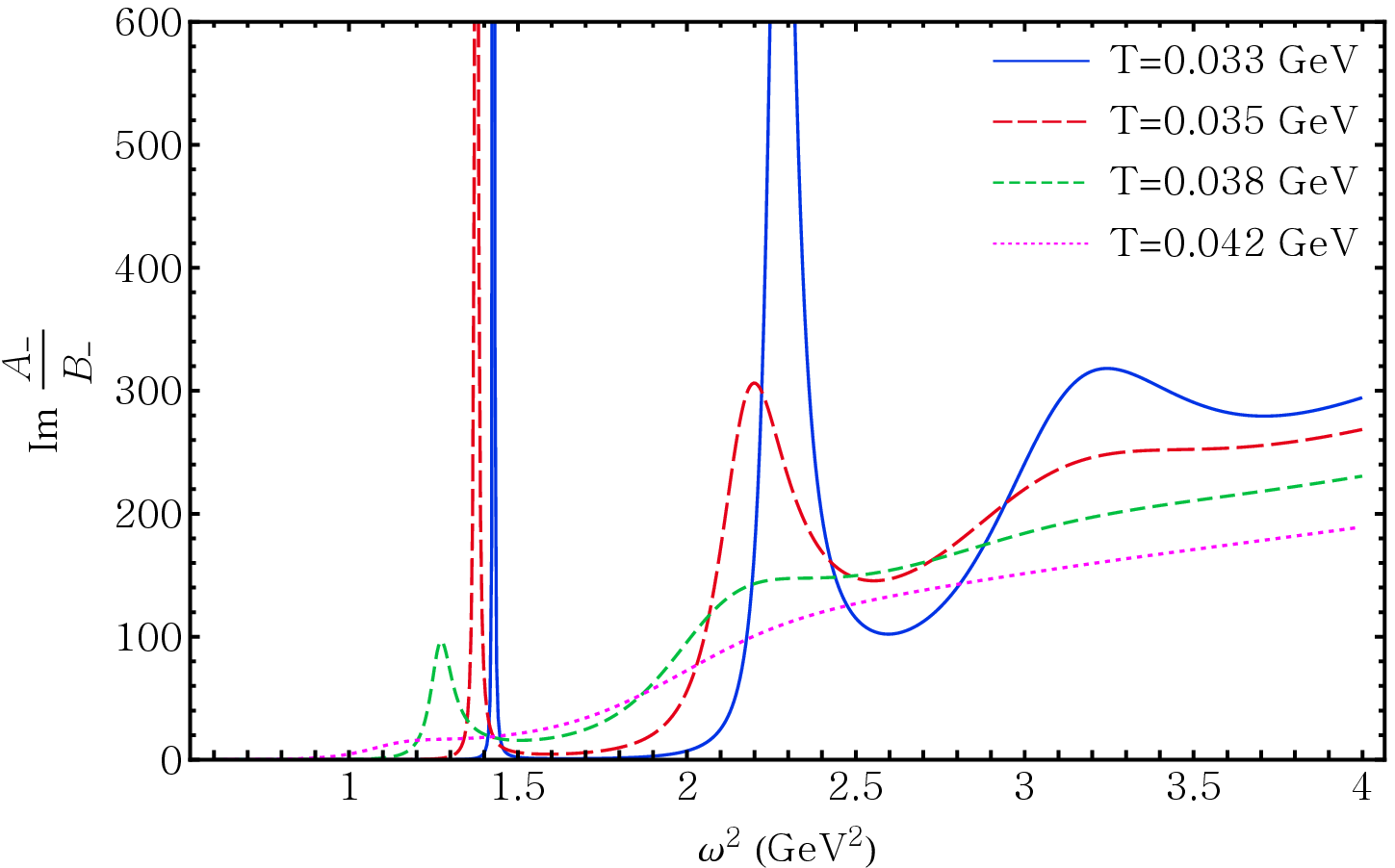}
}
\caption{Thermal spectral functions for a) vector and b) axial-vector mesons.} 
\label{vector_spectra_t}
\end{figure}

Following the steps outlined in Section \ref{spectralmethod} we find the mass spectra for the three meson sectors, incorporating the improved dilaton, dynamical chiral fields, and $m_q=9.75$ MeV. The scalar meson spectrum is shown in Figure \ref{scalar_spectra_t}, and the vector and axial-vector spectra are shown in Figure \ref{vector_spectra_t}. Full melting of the ground state occurs near a temperature of $60$ MeV for the scalar meson and $45$ MeV for the vector and axial-vector mesons.

\begin{figure}[ht] 
\centering
  \includegraphics[width=0.5\textwidth]{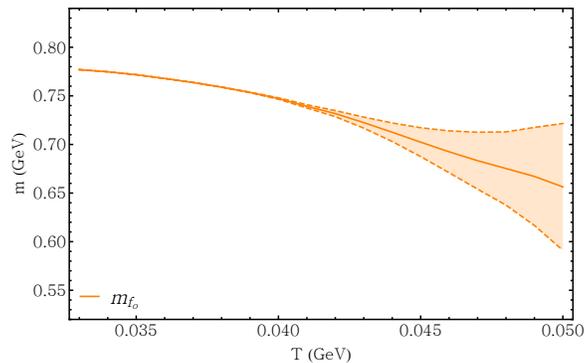}
\caption{Temperature dependence of the ground state scalar mass and peak width. The shaded region corresponds to $m \pm \frac{\Gamma}{2}$.}
\label{scalarmass1_t}
\end{figure}

\begin{figure}[ht] 
\centering
\subfloat[]{
  \includegraphics[width=0.5\textwidth]{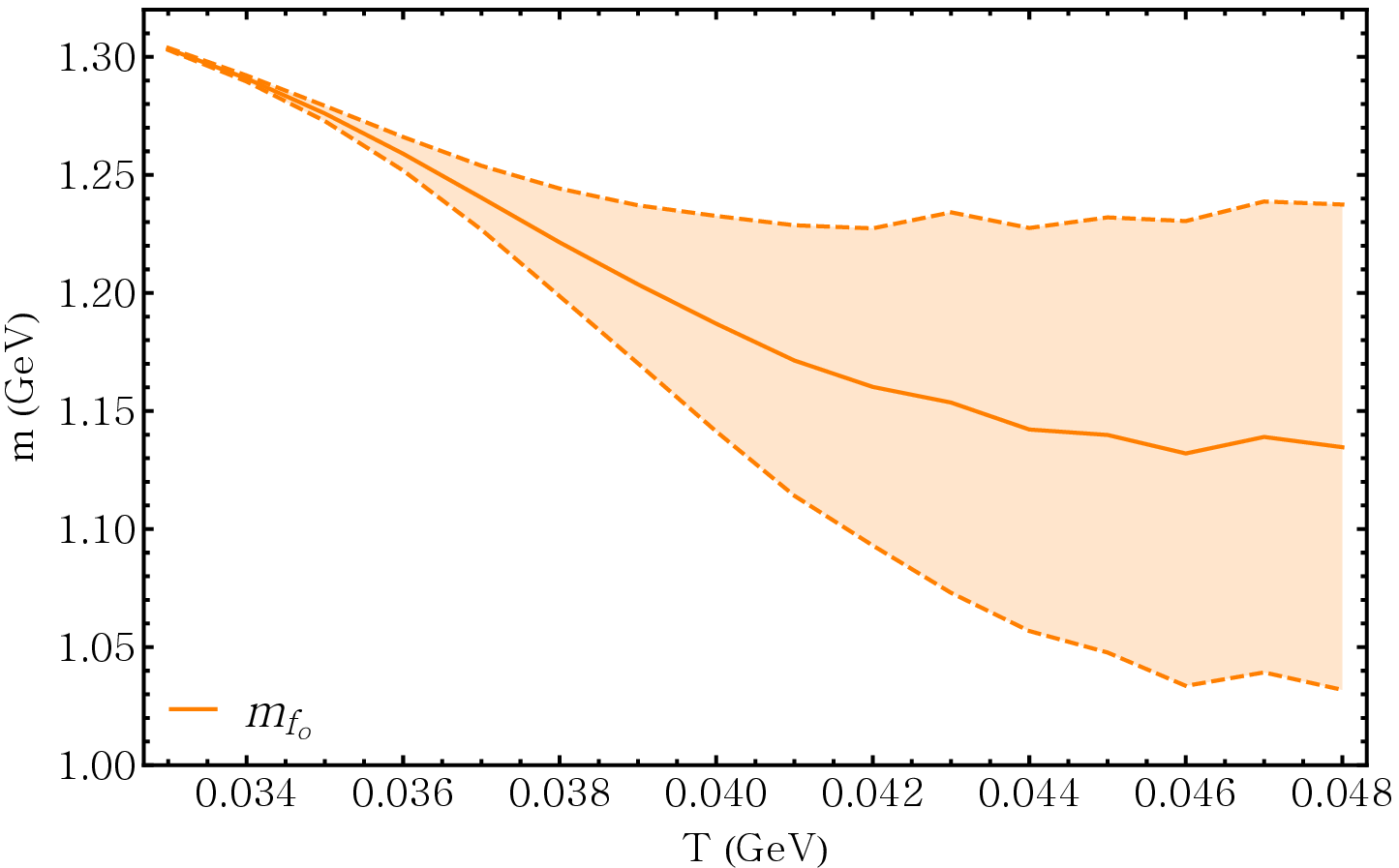}
}
\subfloat[]{
  \includegraphics[width=0.5\textwidth]{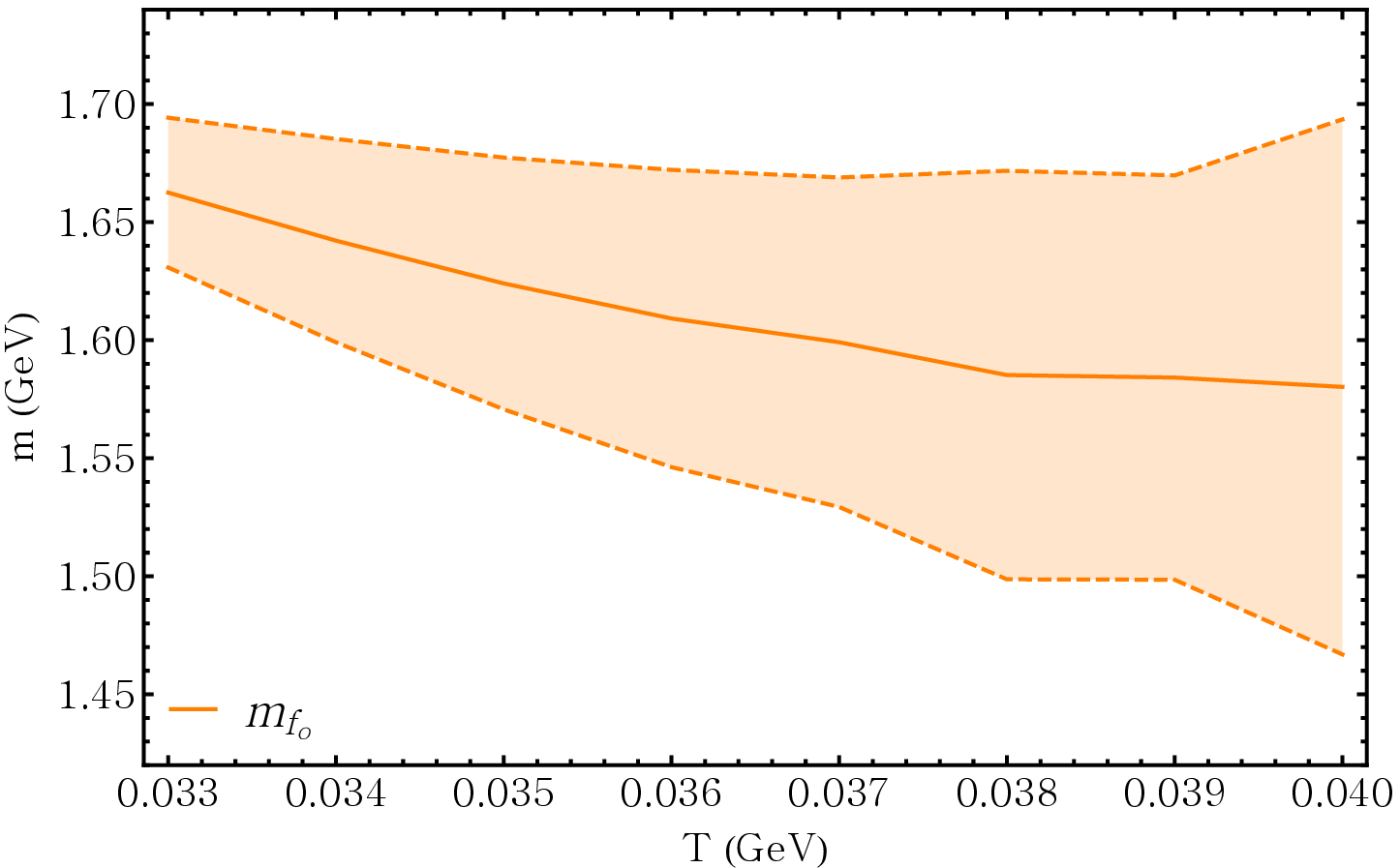}
}
\caption{Temperature dependence of the scalar mass and peak width for (a) the first excited state and (b) the second excited state.}
\label{scalarmass2_t}
\end{figure}

\begin{figure}[htb] 
\centering
  \includegraphics[width=0.5\textwidth]{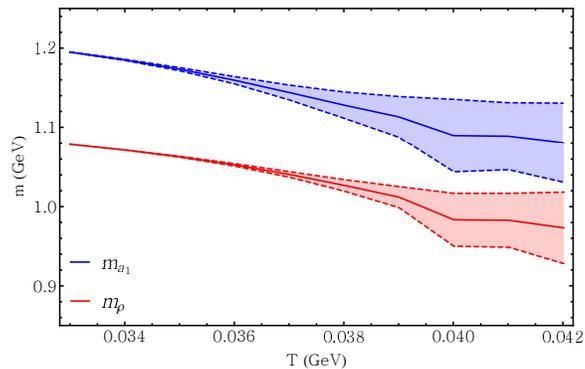}
\caption{Temperature dependence of the ground state vector and axial-vector masses. The mass splitting between axial and vector mesons is evident. The shaded region corresponds to $m \pm \frac{\Gamma}{2}$.} 
\label{vectormass1_t}
\end{figure}

\begin{figure}[htb] 
\centering
\subfloat[]{
  \includegraphics[width=0.5\textwidth]{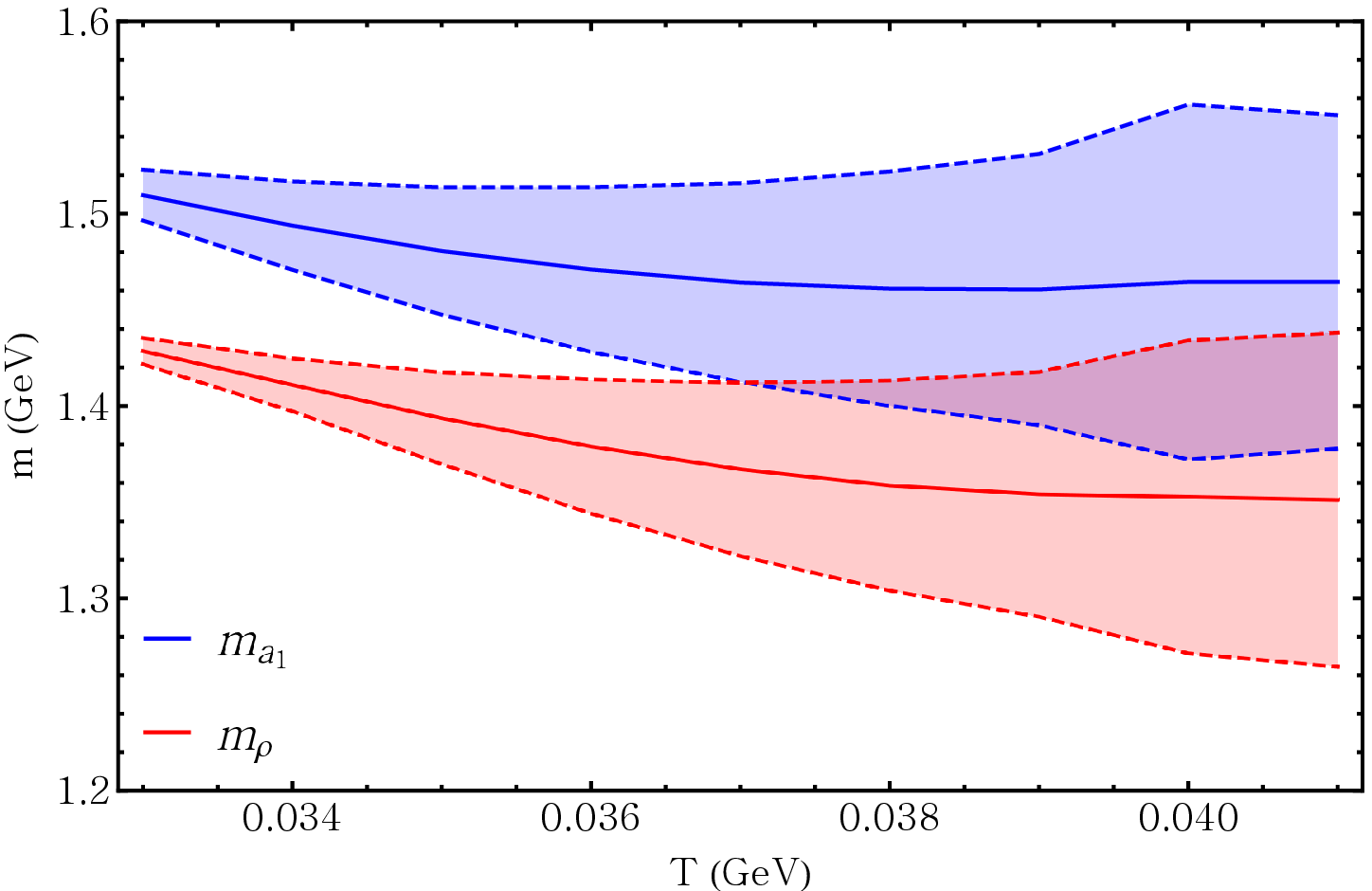}
}
\subfloat[]{
  \includegraphics[width=0.5\textwidth]{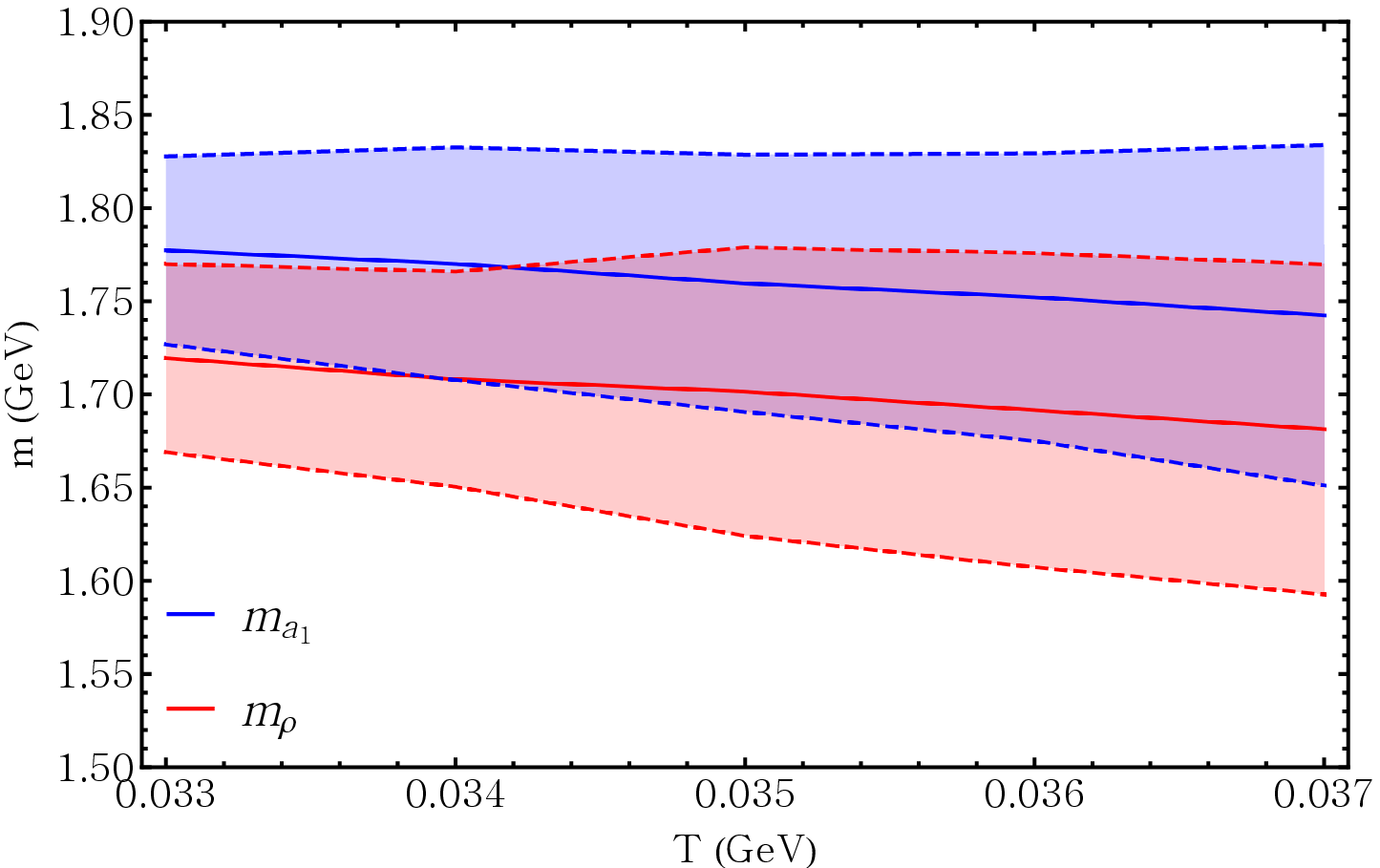}
}
\caption{Temperature dependence of the vector and axial meson masses, for (a) the first excited states and (b) second excited states. Due to the broadening of the peaks, there is larger overlap in the second excited states, but the central values remain evenly spaced.} 
\label{vectormass2_t}
\end{figure}

As the temperature increases, the location of each peak decreases slightly, while the width broadens. For a given temperature we exclude the background and fit each peak numerically with a Breit-Wigner form
\be 
\rho(\omega^2)=\frac{a\ m\ \Gamma\ \omega^b}{(\omega^2-m^2)^2+m^2\Gamma^2}, \label{Breit-Wigner}
\ee
where $a$ and $b$ are numerical parameters, $m^2$ corresponds to the position of the spectral function peaks, and $\Gamma$ is the full-width at half-max of the peaks. 
Figures \ref{scalarmass1_t}, \ref{scalarmass2_t}, \ref{vectormass1_t}, and \ref{vectormass2_t} show the temperature dependence of the mass and peak width of the first three peaks for each sector. 
These plots clearly show the downward shifting of the ground state mass, which is difficult to distinguish in the spectral function plots. 


The new parameterization of the dilaton (\ref{expdil}) produces reasonable spectra despite its UV behavior. The effect of the dynamical chiral field is clear in the splitting between the vector and axial-vector masses, but because peak melting occurs at a lower temperature than the chiral transition, the masses do not converge. The constant mass splitting is shown in Figures \ref{vectormass1_t} and  \ref{vectormass2_t}. 

\section{Finite Temperature, Finite Baryon Chemical Potential} \label{finitemu}
The analysis of Section \ref{zeromu} applies to quark matter at zero baryon chemical potential, where matter and antimatter are present in equal amounts. This situation is comparable to the conditions of the early universe, but is not appropriate for heavy ion collisions. The heavy ions have a non-zero net baryon number that is unchanged throughout the interaction. Because of this, the baryon chemical potential can be considered a proxy for the energy density of the quark-gluon plasma.
In AdS/QCD models, finite-temperature and quark chemical potential are incorporated by modifying the black hole metric.

Following  \cite{Chamblin1999ChargedHolography,Park2010DissociationMedium,Colangelo2011HolographyDiagram}, we consider finite baryon chemical potential by changing the metric (\ref{metricGeneric}) to a charged black hole using the  5D AdS--Reissner-Nordstr{\"o}m metric
\be
f(z)=1-(1+Q^2)\left(\frac{z}{z_h}\right)^4+Q^2\left(\frac{z}{z_h}\right)^6, \label{metricMu}
\ee
where $Q=qz_h^3$ and $0<Q^2<2$. The baryon chemical potential is determined from the charge $q$ of the black hole, along with the position of the horizon. The temperature expression is also modified
\begin{eqnarray}
\mu&=&\kappa \frac{Q}{z_h} \\
T&=&\frac{1}{\pi z_h}\left(1-\frac{Q^2}{2}\right).
\end{eqnarray} 
As in \cite{Colangelo2012TemperatureStudy} we take $\kappa=1$. Any combination of temperature and chemical potential uniquely determines a combination of $z_h$ and $q$. Because both physical quantities depend on the horizon location, it is not possible to explicitly take the temperature to zero, so we examine the limit $T\rightarrow 0$.
We neglect the higher-order corrections to the UV solutions (\ref{scalar_UV}-\ref{vector_UV}) due to the AdS-RN metric, as they do not affect the numerical method. The in-falling solution at the black-hole boundary (\ref{infalling}) becomes \cite{Colangelo2009HolographicMesons}
\be 
\psi_-=(1-u)^{- i\frac{\omega z_h}{2(2-Q^2)}}.
\ee
With this modification, we repeat the analysis of Section \ref{zeromu}.

\subsection{Chiral Dynamics}
\begin{figure}[htb] 
\centering
\subfloat[]{
  \includegraphics[width=0.5\textwidth]{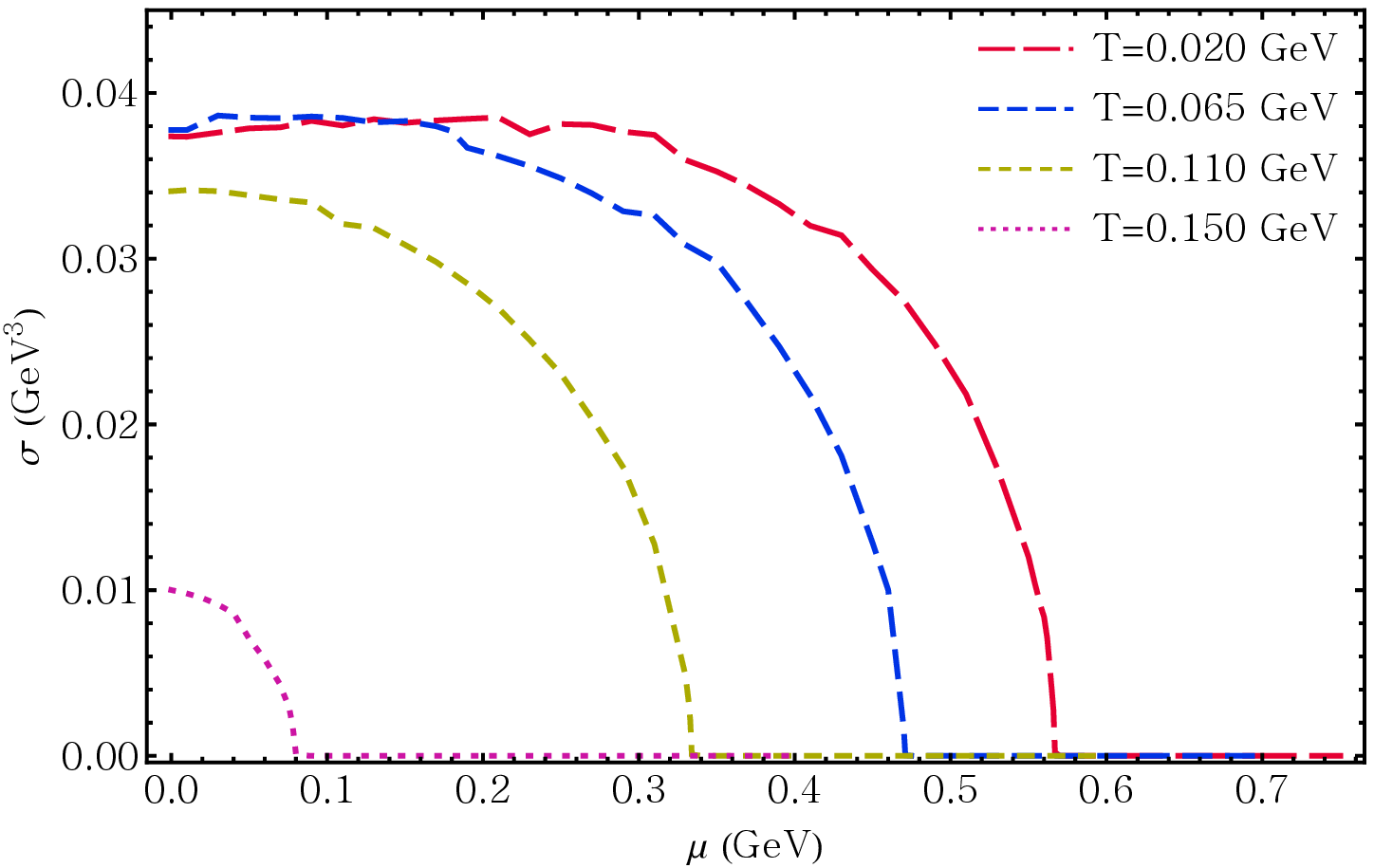}
}
\subfloat[]{
  \includegraphics[width=0.5\textwidth]{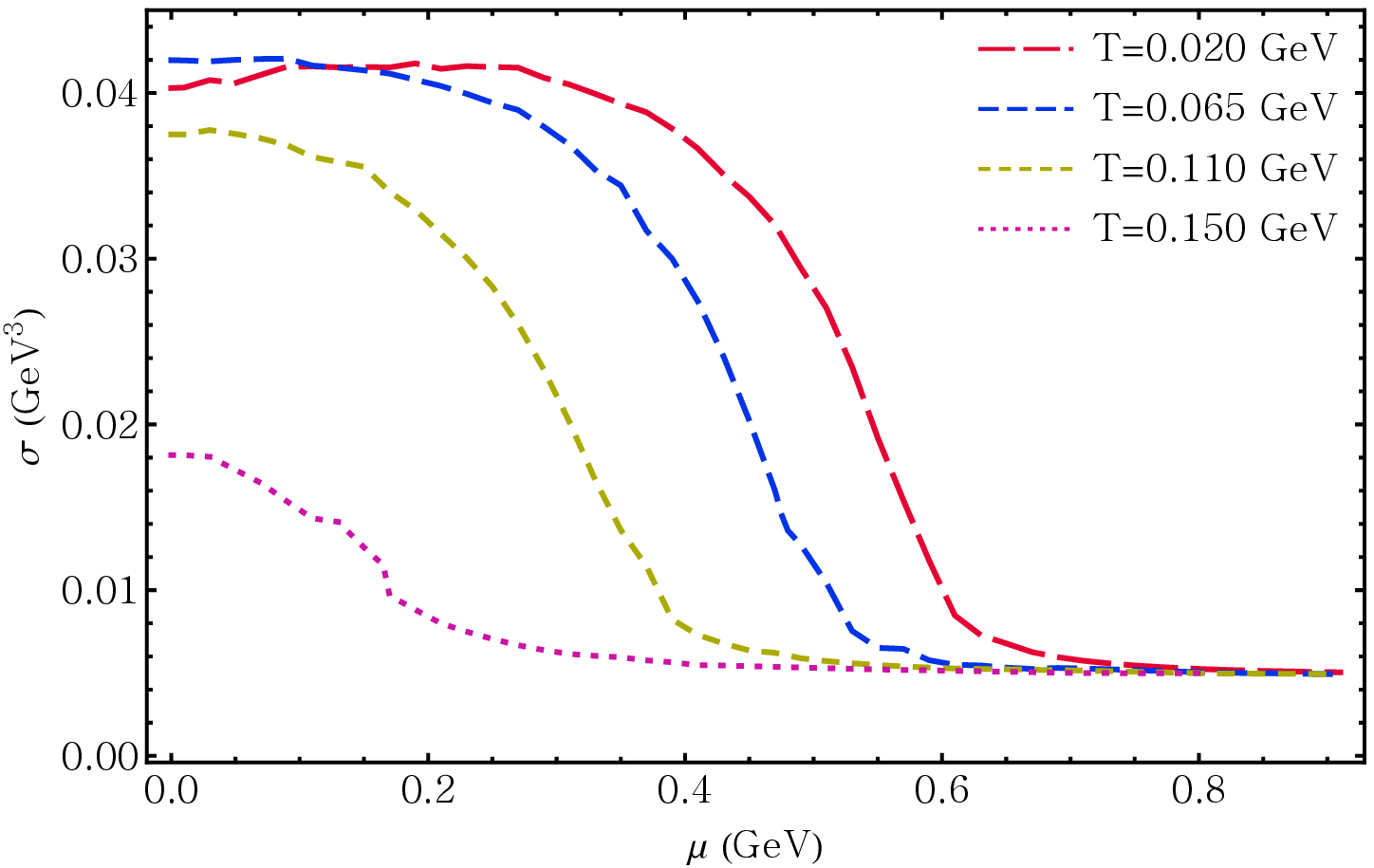}
}
\caption{Dependence of $\sigma$ on $\mu$, for different values of $T$ and (a) $m_q=0$ MeV and (b) $m_q=9.75$ MeV. For zero quark mass there is a higher-order phase transition, while for non-zero quark mass the transition is a rapid crossover.} 
\label{sigma_mu}
\end{figure}

\begin{figure}[!ht] 
\centering
\subfloat[]{
  \includegraphics[width=0.5\textwidth]{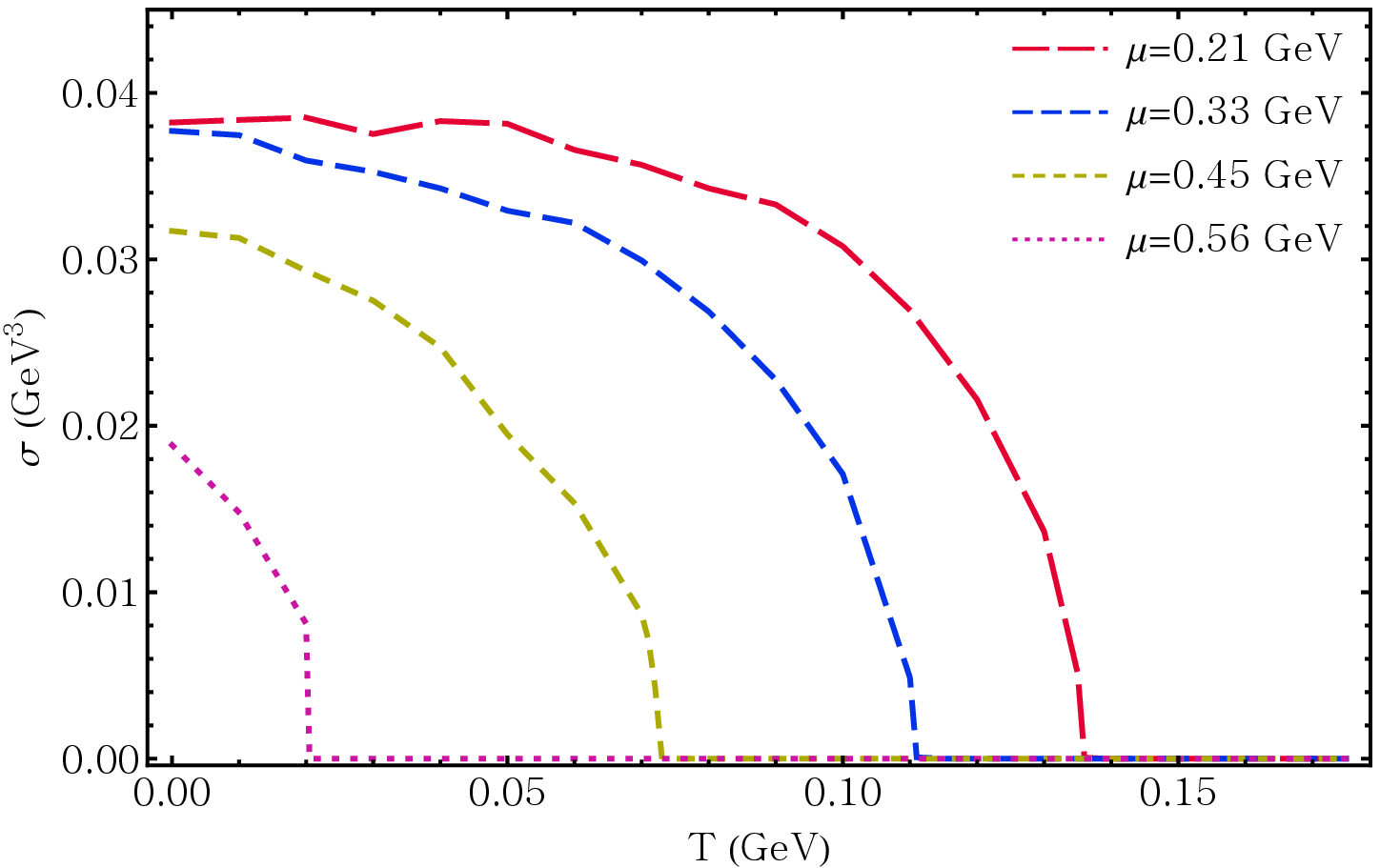}
}
\subfloat[]{
  \includegraphics[width=0.5\textwidth]{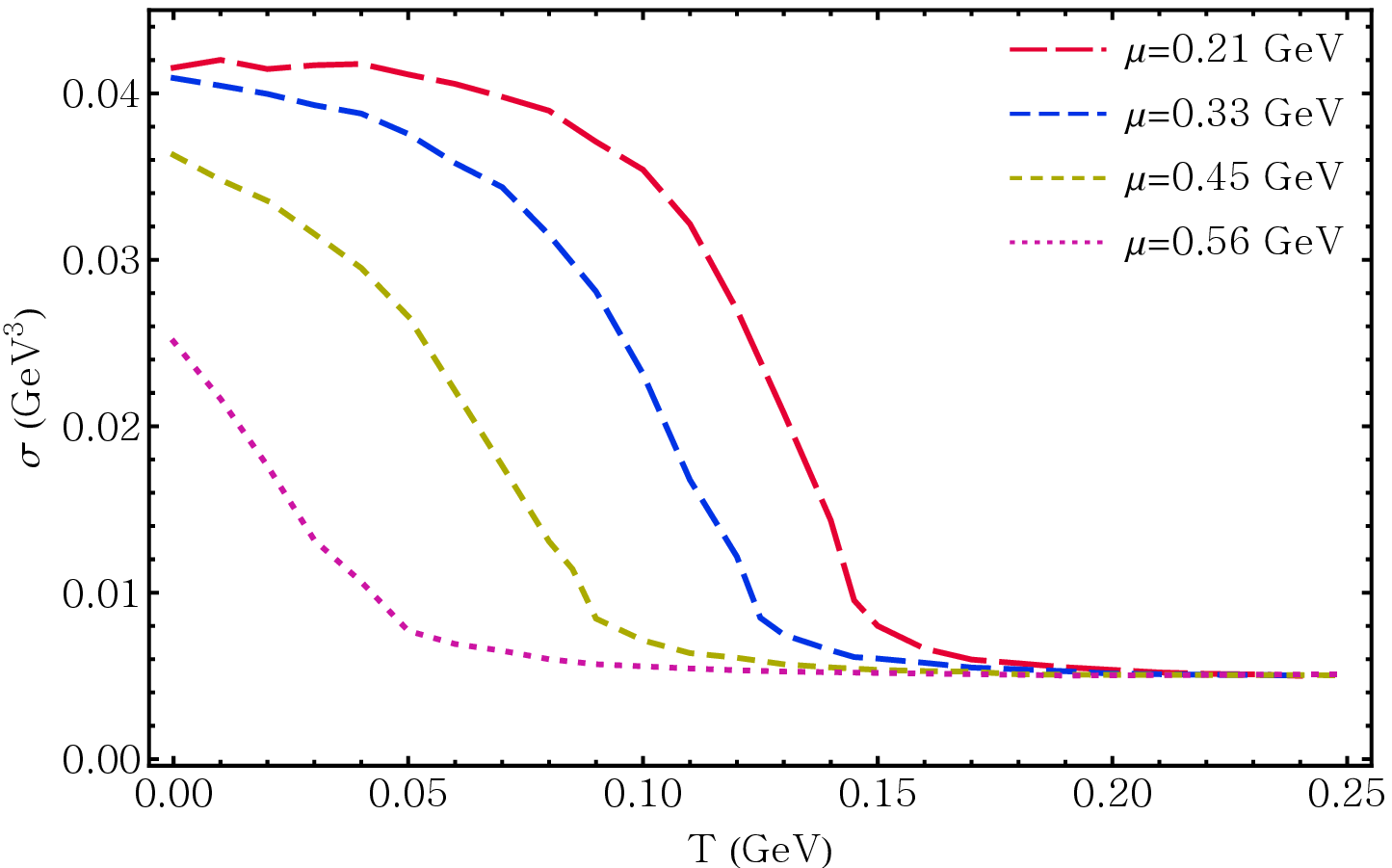}
}
\caption{Dependence of $\sigma$ on $T$, for different values of $\mu$ and (a) $m_q=0$ MeV and (b) $m_q=9.75$ MeV. For higher baryon densities, the chiral phase transition happens at a lower temperature.} 
\label{sigma_t}
\end{figure}

\begin{figure}[!ht]
  \centering
  \includegraphics[width=0.5\textwidth]{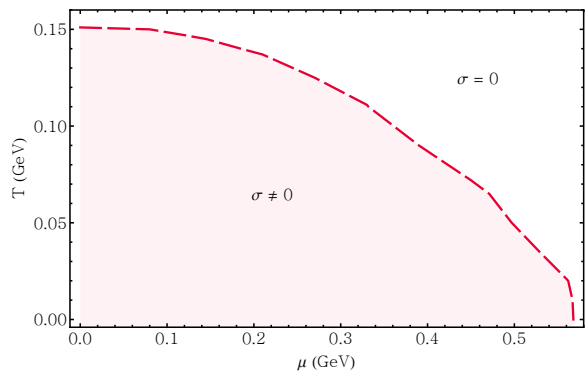}
  \caption{Values of $T$ and $\mu$ for which the quark condensate $\sigma$ goes to zero. The transition is second-order, corresponding to the chiral limit $m_q=0$.}
  \label{muvst}
\end{figure}

We determine the chiral field using the  procedure detailed in Section \ref{Chiral} with the metric (\ref{metricMu}). The boundary conditions of the chiral field are unchanged.
Figures \ref{sigma_mu} and \ref{sigma_t} show the quark chemical potential and temperature dependence of the chiral condensate. The results are qualitatively similar to those obtained for $\mu=0$. When $m_q=0$, the chiral phase transition is second order, while $m_q=9.75$ MeV produces a crossover behavior. Figure \ref{muvst} shows the line in the $T-\mu$ plane where the second-order transition occurs in the chiral limit.

In the chiral limit, we fit the temperature- and chemical potential-dependence of the chiral condensate to the following form
\be 
\frac{\sigma(T)}{\sigma_0}=1-\left(\frac{T}{T_c}\right)^{\alpha},\ \quad   \frac{\sigma(\mu)}{\sigma_0}=1-\left(\frac{\mu}{\mu_c}\right)^{\beta},
\ee
which treats the phase transitions as critical phenomena. For zero baryon chemical potential, we find a critical temperature of $T_c=155$ MeV, and a critical baryon chemical potential of $\mu_c=566$ MeV in the zero-temperature limit. A numerical fitting procedure determines the critical exponents $\alpha=7.3 \pm 0.9$ and $\beta=7.8 \pm 0.6$, suggesting universal critical behavior across the finite-temperature and finite-density regimes.
When the quark mass is finite, the pseudo-critical temperature and chemical potential correspond to the maxima of the susceptibilities $\partial\sigma/\partial T$ and $\partial\sigma/\partial \mu$. Upon analysis, we find that for $m_q=9.75$ MeV the transition occurs at $T_c=151$ MeV and $\mu_c=559$ MeV, similar to the values found in the chiral limit. 
The transition temperature at physical quark mass is consistent with lattice results from various collaborations, which find values around 150 MeV \cite{Aoki2006TheLimit,Aoki2009TheII,Bazavov2012ChiralTransition}. As temperature (chemical potential) increases, the chiral phase transition occurs at a lower chemical potential (temperature). 

Finally, the zero-temperature, zero-density values of the chiral condensate are $\sigma_0=(342.8 \, \mathrm{MeV})^3$ for physical quark mass, and $\sigma_0=(334.4 \,\mathrm{MeV})^3$  in the chiral limit. These values are also consistent with lattice calculations.

\subsection{Meson Melting}
\begin{figure}[!ht]
  \centering
  \includegraphics[width=0.5\textwidth]{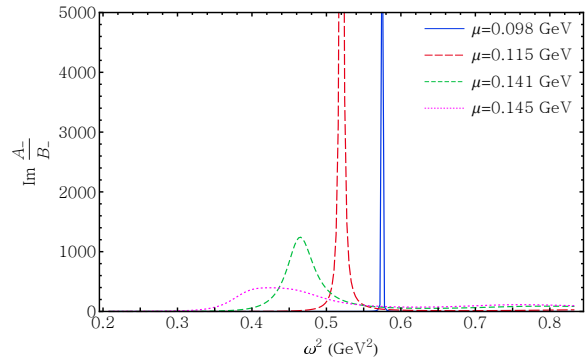}
  \caption{Ground state spectral function for scalar meson at $T=0.02$ GeV and various densities.}
  \label{scalar_spectra_mu}
\end{figure}

\begin{figure}[!ht] 
\centering
\subfloat[]{
  \includegraphics[width=0.5\textwidth]{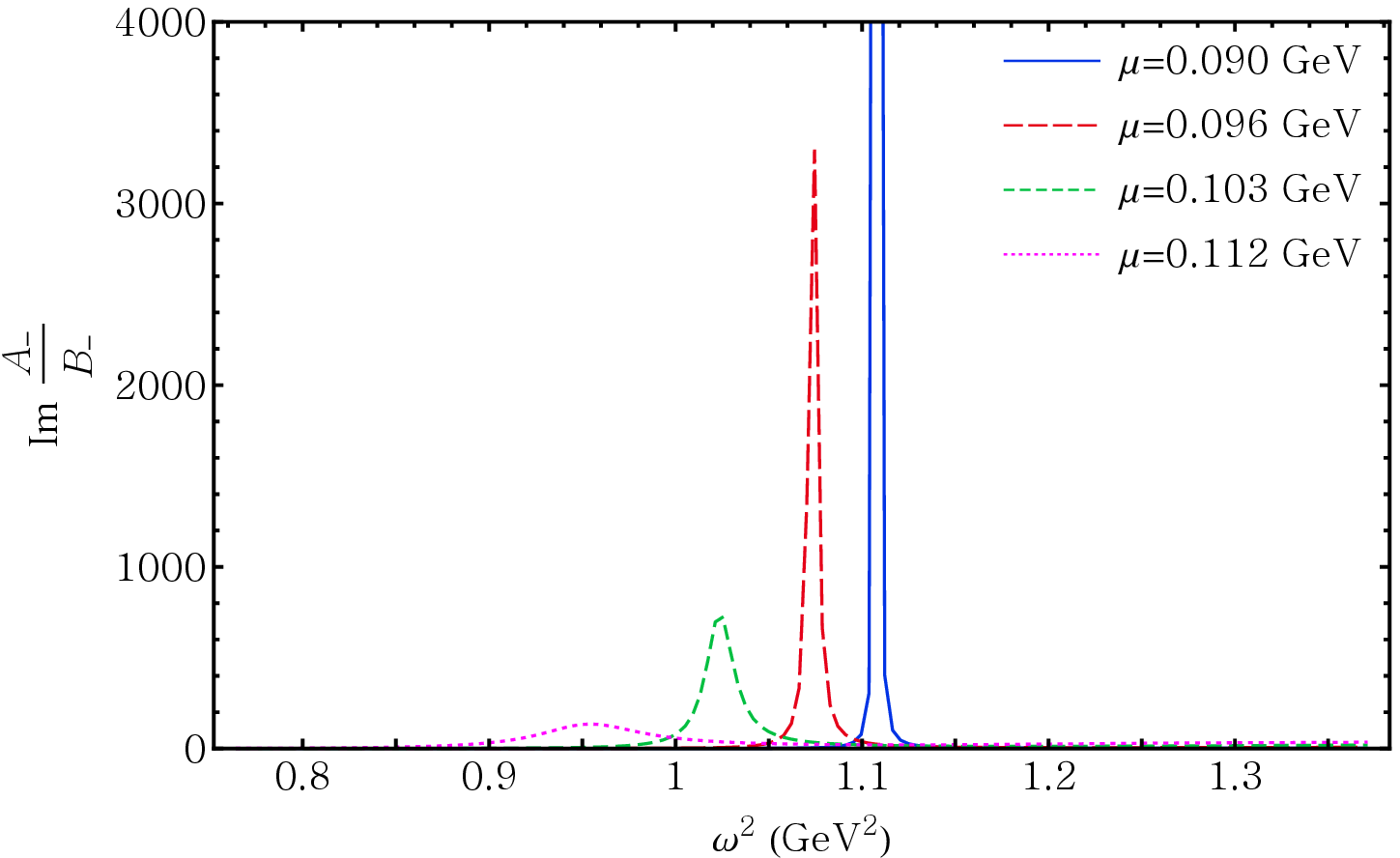}
}
\subfloat[]{
  \includegraphics[width=0.5\textwidth]{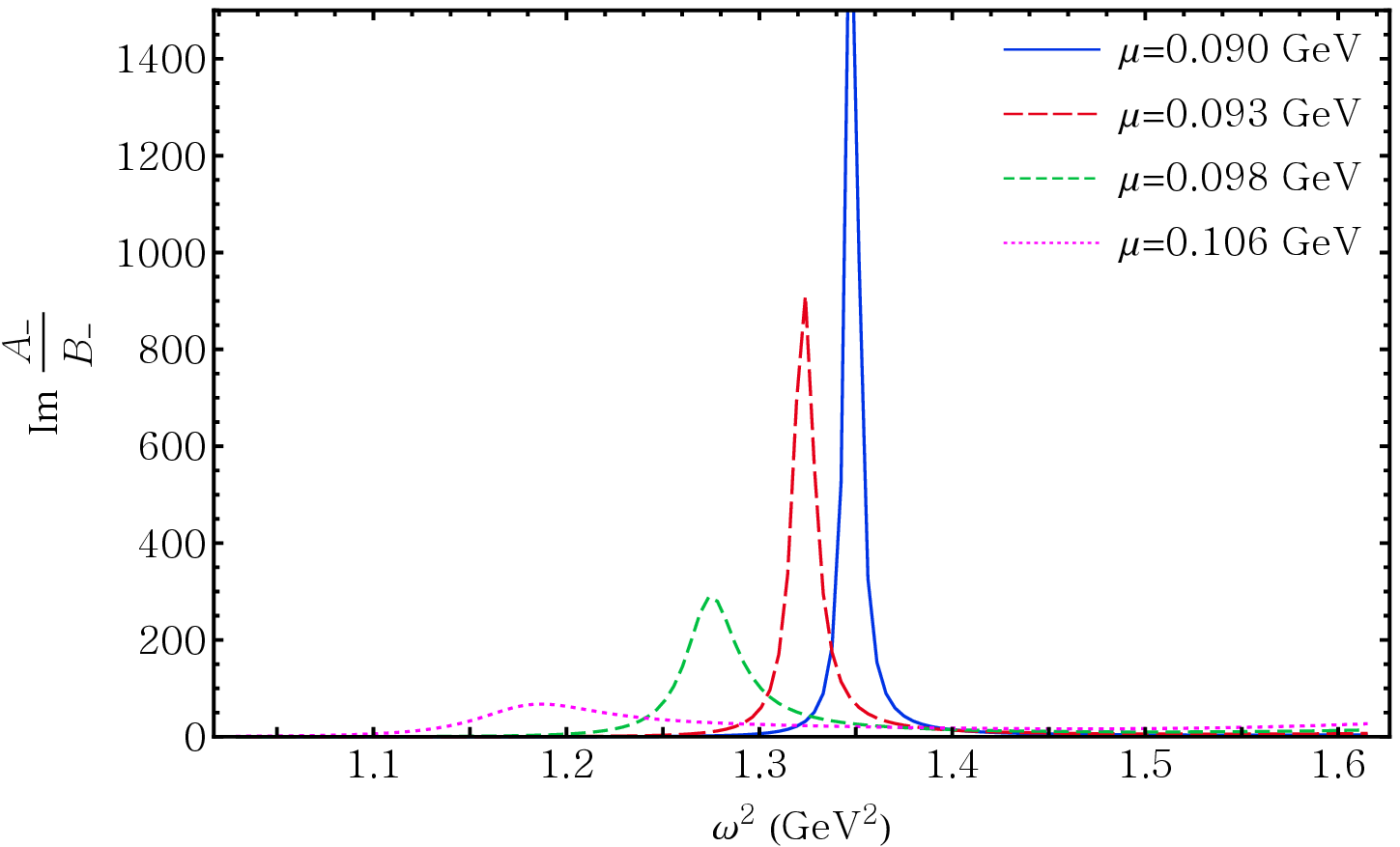}
}
\caption{Ground state spectral function for (a) vector and (b) axial-vector mesons at $T=0.02$ GeV and various densities.} \label{vector_spectra_mu}
\end{figure}

The finite density spectra are qualitatively similar to the finite temperature spectra. We employ the numerical method outlined in Section \ref{spectralmethod} and Appendix \ref{numericalmethod}, with a low temperature of $T=20$ MeV and various values of the chemical potential. 
In the case of zero baryon density, using the AdS-Schwarzschild metric, the background behavior of the mass spectra is monotonically increasing and easily removed prior to Breit-Wigner fitting. 
However, modifying the metric to AdS-RN causes significant oscillations in the spectral function background, which cannot be adequately removed. For this reason, we plot only the ground state peaks. 
Figures \ref{scalar_spectra_mu} and \ref{vector_spectra_mu} show the melting of the ground state mass peaks for the three sectors. Melting of the scalar ground state occurs near a chemical potential of 150 MeV, and melting of the vector and axial ground states occurs around 115 MeV.

\begin{figure}[!ht] 
\centering
\subfloat[]{
  \includegraphics[width=0.5\textwidth]{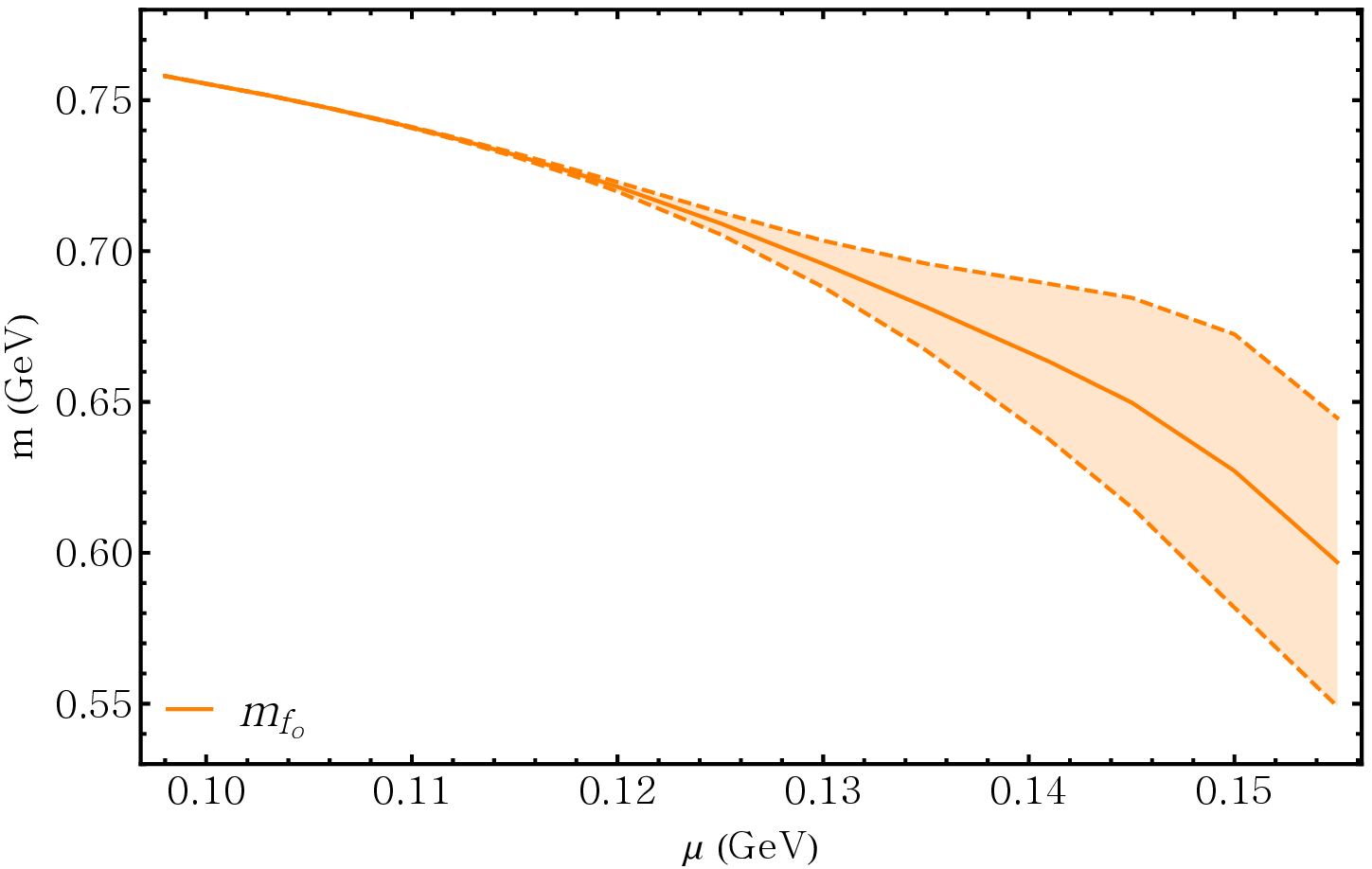}
}
\subfloat[]{
  \includegraphics[width=0.5\textwidth]{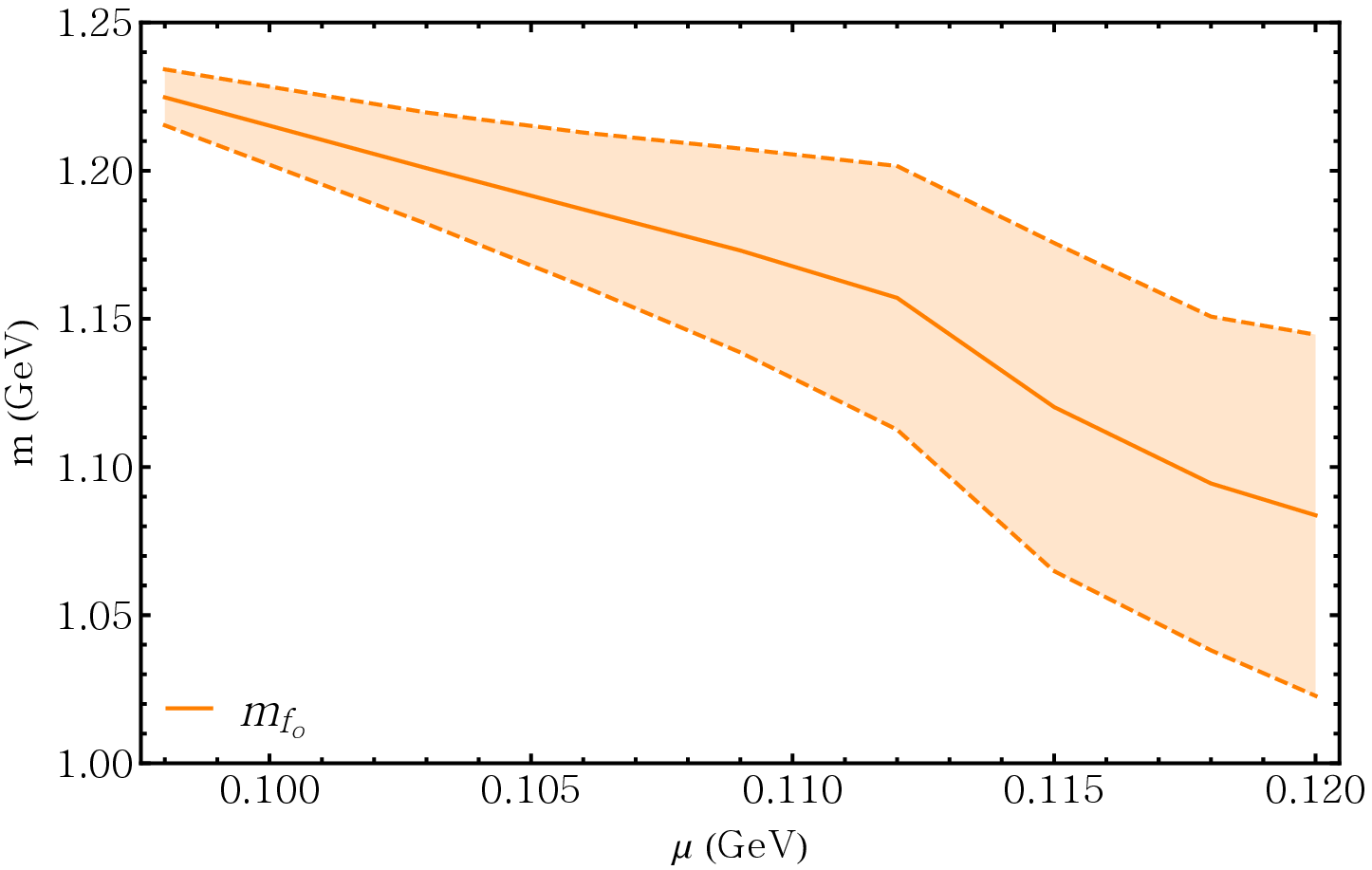}
}
\caption{Baryon chemical potential dependence of the scalar meson mass for (a) the ground state and (b) first excited state, with $T=0.02$ GeV. The shaded region corresponds to $m \pm \frac{\Gamma}{2}$.}
\label{scalarmass_mu}
\end{figure}

\begin{figure}[!ht] 
\centering
\subfloat[]{
  \includegraphics[width=0.5\textwidth]{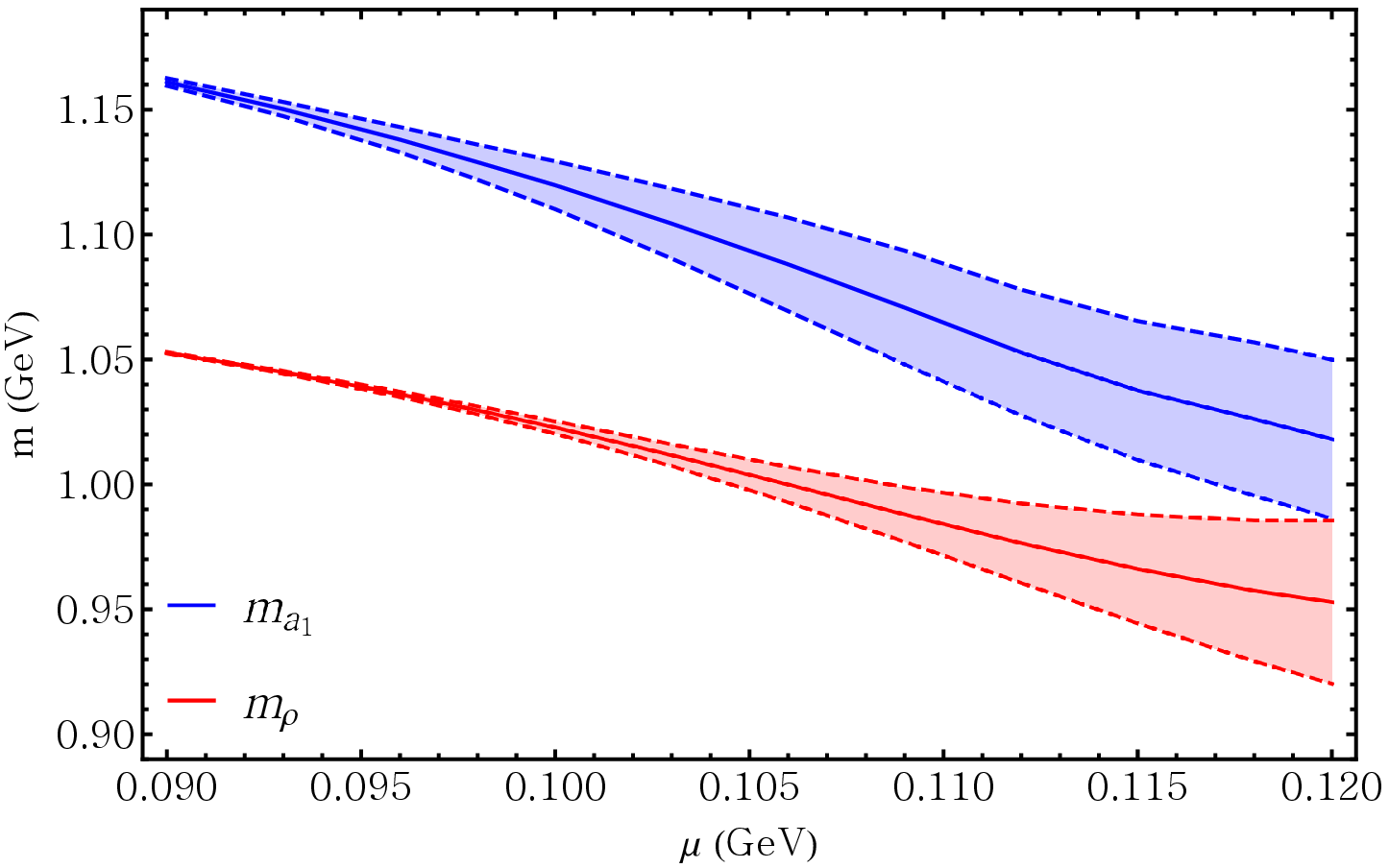}
}
\subfloat[]{
  \includegraphics[width=0.5\textwidth]{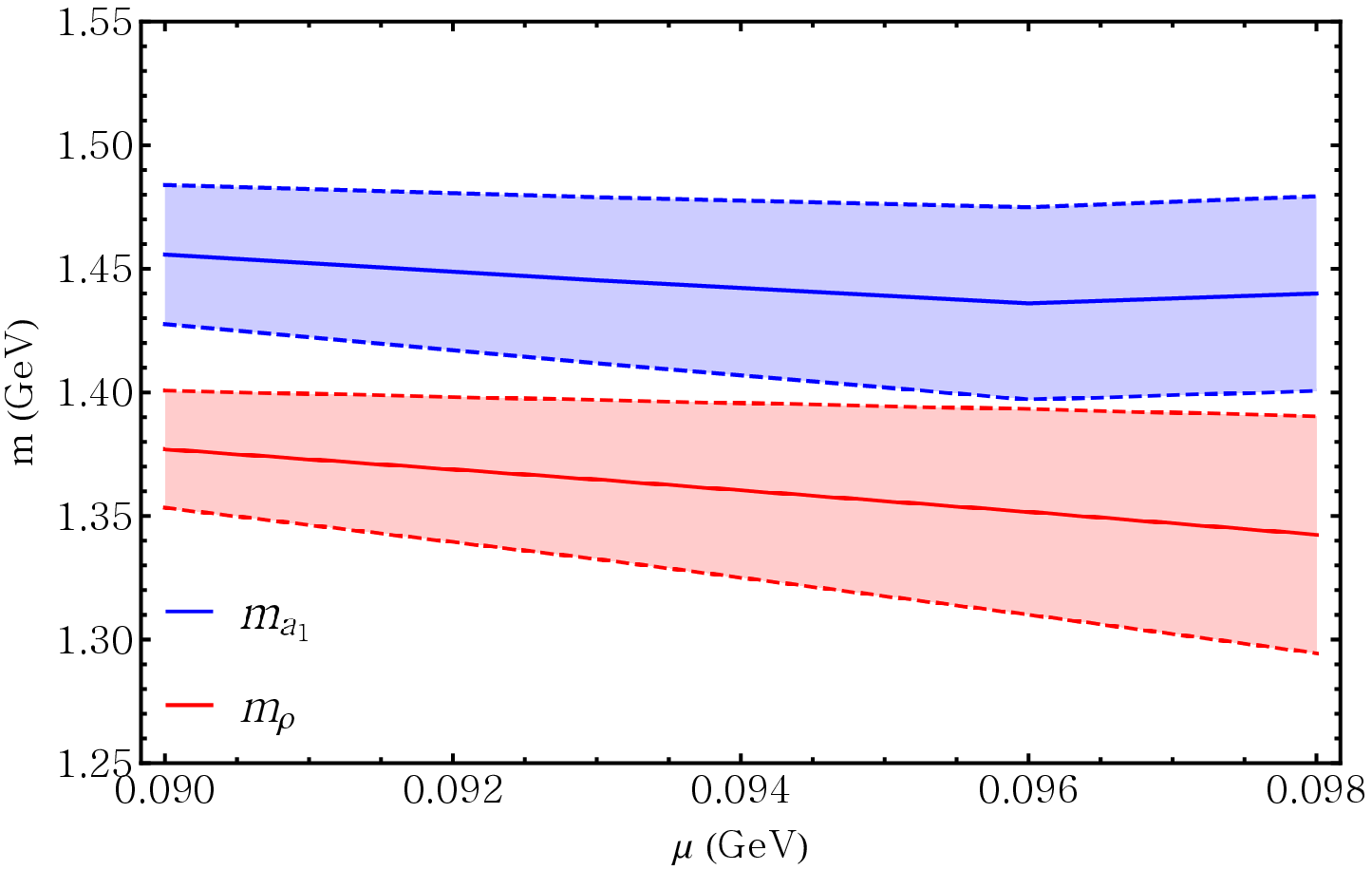}
}
\caption{Baryon chemical potential dependence of (a) the ground state and (b) first excited state vector and axial meson masses, with $T=0.02$ GeV. The mass splitting between axial and vector mesons is clear, with slight convergence of the ground state masses at higher density.}
\label{vectormass_mu}
\end{figure}

As in the finite-temperature case, increasing the quark chemical potential causes the masses to decrease while the peaks broaden. Numerically fitting the ground and first excited states to the Breit-Wigner form (\ref{Breit-Wigner}) allows for a more precise analysis. Figures \ref{scalarmass_mu} and \ref{vectormass_mu} show, for $T=20$ MeV, the quark density dependence of the mass and peak width of each sector by showing the region bounded by $m \pm \Gamma/2$. It is worth noting that the ground state vector and axial-vector masses appear to converge as the baryon density increases. However, the mass splitting for the first excited state remains roughly constant over the small range plotted. As in Section \ref{zeromu} we find that meson melting occurs well before the chiral phase transition. 


\section{Conclusions}
In this work, we investigate the chiral phase transition in soft-wall AdS/QCD at finite temperature and baryon chemical potential.
Higher-order terms in the scalar potential are included in the action, allowing for independent sources of explicit and spontaneous chiral symmetry breaking.
We find the equations of motion for scalar, vector, and axial-vector mesons and numerically find the spectral functions for each sector.
Following the precedent of \cite{Chelabi2016ChiralAdS/QCD}, we use a dilaton profile that is negative quadratic in the UV but positive quadratic in the IR in order to find a model that permits a finite chiral field while also achieving reasonable Regge trajectories for the meson spectra.
However, we use a modified dilaton parameterization that improves the resulting meson spectra and allows exploration of the spectra at lower temperatures.

All analysis is performed in both the finite temperature (AdS-Schwarzschild) and finite chemical potential (AdS--Reissner-Nordstr{\"o}m) regimes.
We numerically determine the chiral field for use in the meson equations of motion and to find the chiral transition which behaves similarly in the two regimes. 
In the chiral limit, the transition is second-order, with critical exponents that suggest universal critical behavior.
When a physical quark mass is implemented, the transition becomes a rapid crossover.

The meson spectral functions show the melting of the bound states as temperature or density is increased. 
The central value of the peaks shifts downward slightly as the peaks broaden. Eventually all states melt, beginning with the excited states, until the ground states melt at a temperature  about 40-60 MeV in the zero-chemical potential case.
The behavior is qualitatively similar when chemical potential is introduced, where the ground states melt at a baryon chemical potential  around 110-150 MeV.
We examine the mass splitting between the vector and axial-vector meson masses, a sign of chiral symmetry breaking.
The splitting remains roughly constant at all values of $T$ and $\mu$, which is sensible given that the bound state melting occurs before the chiral phase transition.

This paper presents improvements upon earlier results in this area by incorporating realistic chiral symmetry breaking in the finite chemical potential regime and including quantitative exploration of the meson spectra.
As in other holographic models of meson melting, we find a melting temperature that is below the deconfinement temperature predicted by lattice QCD. For a full analysis of the deconfinement phase transition, further study should focus on the Hawking-Page transition between the AdS-BH and a thermal AdS solution. 
Meson spectra results could possibly be further improved with a different parameterization of the dilaton. 
In addition, the dilaton and metric for this model are treated as background, rather being solved dynamically from the gravity action.

\section*{Acknowledgments}
The authors would like to thank Macalester College for funding this work.


\appendix
\section{Numerical Method for Finding Spectral Functions} \label{numericalmethod}
We follow a numerical procedure to find the spectral function \cite{Miranda2009Black-holeModel,Teaney2006FiniteTheory}
We write (\ref{outgo}, \ref{infall}) in matrix form
\be 
\left(\begin{matrix}
\psi_- \\
\psi_+
\end{matrix} \right)
=
\left(
\begin{matrix}
A_- & B_- \\
A_+ & B_+ 
\end{matrix} \right)
\left(\begin{matrix}
\psi_1 \\
\psi_2
\end{matrix}\right), \label{matrixHorizon}
\ee
and define a similar matrix to perform the inverse operation
\be 
\left(\begin{matrix}
\psi_1 \\
\psi_2
\end{matrix} \right)
=
\left(
\begin{matrix}
C_1 & D_1 \\
C_2 & D_2 
\end{matrix} \right)
\left(\begin{matrix}
\psi_- \\
\psi_+
\end{matrix}\right). \label{matrixUV}
\ee 
The allowed values of $\omega$ are those that allow the physical boundary conditions to be matched in both the UV and near-horizon limits.
That is, the in-falling solution $\psi_-$ near the horizon will be composed primarily of the Bogoliubov-normalizable UV wavefunction $\psi_1$.
This is equivalent to finding the values of $\omega$ that cause the matrices in (\ref{matrixHorizon}, \ref{matrixUV}) to be approximately diagonal.
In \cite{Miranda2009Black-holeModel}, it was shown that the retarded Green's function is proportional to the imaginary part of the ratio $A_-/B_-$.
In order to find this ratio for a given value of $\omega$, we use the following procedure. 
It is evident that
\be 
\left(
\begin{matrix}
C_1 & D_1 \\
C_2 & D_2 
\end{matrix} \right)^{-1} =
\left(
\begin{matrix}
A_- & B_- \\
A_+ & B_+ 
\end{matrix} \right). \label{vectordef}
\ee
Performing the matrix inversion and solving for the combination of interest gives 
\be 
\frac{A_-}{B_-}=-\frac{D_2}{D_1}.
\ee

We now use  (\ref{vectordef}) to construct vectors that incorporate the function and its derivative 
\be 
\left(\begin{matrix}
\psi_j(z) \\
\partial_z\psi_j(z)
\end{matrix} \right)
=
\left(
\begin{matrix}
\psi_-(z) & \psi_+(z) \\
\partial_z \psi_-(z) & \partial_z \psi_+(z)
\end{matrix} \right)
\left(\begin{matrix}
C_j \\
D_j
\end{matrix}\right), \qquad (j=1,2). \label{invertme}
\ee
For each value of $\omega,$ we numerically integrate $\psi_1,\psi_2$ from $z\sim 0$ to $z\sim z_h$ and use the values in the expression found by inverting (\ref{invertme})
\be 
\frac{A_-}{B_-}=-\frac{D_2}{D_1}=\frac{\partial_z\psi_-(z_h)\psi_2(z_h)-\psi_-(z_h) \partial_z \psi_2(z_h)}{\partial_z\psi_-(z_h)\psi_1(z_h)-\psi_-(z_h) \partial_z \psi_1(z_h)}.
\ee
The imaginary part of this ratio is proportional to the retarded two-point Green's function (\ref{Green}). This is the numerical quantity that is referred to as the spectral function in this work. It is important to note that $z=0$ and $z=z_h$ are singular points in the equations of motion, so we numerically integrate at points close to these limits.  

\bibliographystyle{utphys.bst}
\bibliography{SlashTheoryEprint.bib}

\end{document}